\documentclass[journal]{IEEEtran}

\usepackage{amsthm, amsmath, amsfonts, amssymb, amscd}
\usepackage[dvips]{graphicx}
\usepackage{epsfig}
\usepackage{epstopdf}
\usepackage{graphicx}
\usepackage{subfigure}
\usepackage{enumerate}
\usepackage{color}
\usepackage{cite}
\usepackage{tikz}
\usepackage{lipsum}
\usepackage{siunitx}
\usepackage{tabularx}
\usepackage{textcomp}
\usepackage{stfloats}
\usepackage{balance}

\usepackage[english]{babel}

\usepackage{caption}
\newtheorem{theorem}{Theorem}

\newcommand{\ts}{T_s}
\newcommand{\pavg}{P_{\text{\footnotesize avg}}}
\newcommand{\ptx}{P_{\text{\footnotesize T}}}
\newcommand{\tauk}{\tau_{\text{\footnotesize k}}}
\newcommand{\taukf}{\tau_{\text{\footnotesize k}}^{\text{\footnotesize (f)}}}
\newcommand{\taukr}{\tau_{\text{\footnotesize k}}^{\text{\footnotesize (r)}}}
\newcommand{\alphakf}{\alpha_{\text{\footnotesize k}}^{\text{\footnotesize (f)}}}
\newcommand{\alphak}{\alpha_{\text{\footnotesize k}}}
\newcommand{\alphakr}{\alpha_{\text{\footnotesize k}}^{\text{\footnotesize (r)}}}
\newcommand{\tauj}{\tau_{\text{\footnotesize j}}}

\newcommand{\phil}{\varphi_{\text{\footnotesize l}}}
\newcommand{\rhok}{\rho_{\text{\footnotesize k}}}
\newcommand{\rhot}{\rho_{\text{\tiny T}}}
\newcommand{\rhokf}{\rho_{\text{\footnotesize k}}^{\text{\footnotesize (f)}}}
\newcommand{\rhokr}{\rho_{\text{\footnotesize k}}^{\text{\footnotesize (r)}}}

\newcommand{\sigjk}{\sigma_{\text{\tiny j,k}}^2}
\newcommand{\signk}{\sigma_{\text{\tiny n,k}}^2}
\newcommand{\sbtot}{B^{\text{\tiny k,j}}_{\text{\tiny s,tot}}}
\newcommand{\sbr}{B_{\text{\tiny s,R}}}
\newcommand{\sbd}{B_{\text{\tiny s,D}}}
\newcommand{\sba}{B_{\text{\tiny s,A}}}
\newcommand{\sbtotal}{B_{\text{\tiny s,tot}}}
\newcommand{\dsinc}{\text{Dsinc}}
\newcommand{\sinc}{\text{sinc}}
\newcommand{\nrange}{N_{\text{\tiny R,FFT}}}
\newcommand{\ndopp}{N_{\text{\tiny D,FFT}}}
\title{\textcolor{black}{Information Theoretic Bounds and Waveform Optimization for Joint Radar and Communication with Semi-}Passive Targets}
\author{\IEEEauthorblockN{Ganesan Thiagarajan and Sanjeev Gurugopinath}
\thanks{Ganesan Thiagarajan is with MMRFIC Technology Private Limited, Bengaluru 560043, India.}
\thanks{Sanjeev Gurugopinath is with the Department of Electronics and Communication Engineering, PES University, Bengaluru 560085, India.}
\thanks{Emails: gana@mmrfic.com, sanjeevg@pes.edu}}

\begin{document}
\maketitle

\begin{abstract}
\textcolor{black}{In this paper, we derive the information theoretic performance bounds on communication data rates and errors in parameter estimation, for a joint radar and communication (JRC) system. We assume that targets are semi-passive, i.e.~they use active components for signal reception, and passive components to communicate their own information. Specifically, we let the targets to have control over their passive reflectors in order to transmit their own information back to the radar via reflection-based beamforming or backscattering. We derive the Cram\'er-Rao lower bounds (CRBs) for the mean squared error in the estimation of target parameters. The concept of a target ambiguity function (TAF) arises naturally in the derivation CRBs. Using these TAFs as cost function, we propose a waveform optimization technique based on calculus of variations. Further, we derive lower bounds on the data rates for communication on forward and reverse channels, in radar-only and joint radar and communications scenarios. Through numerical examples, we demonstrate the utility of this framework for transmit waveform design, codebook construction, and establishing the corresponding data rate bounds.}
\end{abstract}


\section{Introduction} \label{SecIntro}
\subsection{Background and Motivation} \label{SubSecBM}
The tremendous increase in demand for wireless applications and a sharp rise in the number of connected devices has caused severe spectrum shortage and congestion of existing bands. 
%Among several communication systems with under-utilized spectra, radar systems standout, which have been allotted a large portion of spectral resources. 
Towards this end, research attention on coexistence of other communication systems in radar bands has significantly grown in the past decade or so. For instance, DARPA funded a research project called \emph{shared spectrum access for radar and communications} (SSPARC), for eliciting technologies where both radar and communication can co-exist \cite{DARPA_SSPARC}. Joint radar and communications (JRC) finds its utility in several civilian applications such as millimeter wave communications, WiFi-based localization, unmanned aerial vehicles communications, RFID, etc., and also in military applications. Excellent recent surveys on applications and future research directions on JRC are given in \cite{Paul_Access_2017, Zheng_SPM_2019, Liu_TCom_2020}. As discussed in \cite{Liu_TCom_2020}, there are two main research directions in JRC, namely (i) radar-communication coexistence (RCC), and (ii) dual-functional radar-communication (DFRC) systems  \cite{Paul_Access_2017}, \cite{Chiriyath_TCCN_2017}. The goal in RCC is to design efficient interference management techniques for coexistence of a communication system in the radar band \cite{Zhang_STSP_2014, Leigsnering_STSP_2015, Zheng_STSP_2018, Gui_STSP_2018, Kilani_ComLet_2018, Kumari_2020, Liu_TSP_2020}. On the other hand, the goal in a DFRC system is the joint design of sensing and signaling operations through a single hardware for both radar and communications in applications such as indoor radars \cite{Roehr_MTT_2008, Wang_MTT_2014, ElAbsi_Access_2018} and radars for vehicular networks \cite{Hata_ITS_2016}, \cite{Zochmann_Access_2019}.

Recent radar applications such as automotive radars \cite{Patole_SPM_2017}, \cite{Waldschmidt_JM_2021} and drone tracking \cite{Guvenc_ComMag_2018}, \cite{Dogru_RAL_2020} have introduced new set of challenges and design considerations for spectrum sharing in JRC. Tracking of unmanned aerial vehicles (UAV) and drones finds applications in security and surveillance \cite{Zhang_TASE_2019}. Latest developments in joint data transmission and target parameter estimation, by resource sharing, have added new dimensions to this problem \cite{Guerra_VTM_2020}. In many such applications, it is necessary to identify a UAV as a friend or a foe, to ensure secure communication \cite{Tang_TIFS_2019}. Since UAVs such as drones operate on battery power, passive communication over active is preferred from a UAV to a JRC transceiver \cite{Stockman_IRE_1948}, \cite{Blunt_AES_2010}, for the following two reasons. First, this enables a UAV to hide its presence and identity from adversaries. Second, employing technologies such as opportunistic ambient backscattering communication (ABC) \cite{Kishore_TCCN_2019} and intelligent reflecting surfaces (IRS) \cite{Wang_SPL_2020}, \cite{Zhou_TSP_2020} that require only a small fraction of power of active transmission, enables the battery power of a UAV to be utilized efficiently.

To the best of our knowledge, a study on the fundamental limits on the performance of JRC with passive targets in terms of joint parameter estimation and communication data rates has not been addressed in literature so far. In this case, crucial design trade-offs between the performance of parameter estimation and data transmission, must be derived and studied based on fundamental performance metrics such as mutual information achievable over the JRC communication links and limits on the target parameter estimates, in terms of the Cram\'er-Rao lower bounds (CRB).

\subsection{Related Work}  \label{SubSecRelatedWork}
Studies on information theory for radar started with the seminal works by Woodward \cite{Woodward_IRE_1951, Woodward_PM_1951, Woodward_IEE_1952, Woodard_Book_1953}, and Davies \cite{Davies_IEE_1952}.  Woodward and Davies showed that a receiver which maximizes the a posteriori probability is optimal, and it reduces to a \emph{correlation receiver} in the case of additive white Gaussian noise. In particular, as opposed to the signal-to-noise (SNR) maximization, the goal is to maximize the `a posteriori probability', which improves the detection probability. The Wigner-Ville transformation was proposed as the unnormalized ambiguity function, which was used to analyse the simultaneous range and velocity estimation resolution \cite{Woodard_RRE_1967}. Klauder used this ambiguity function to design better radar waveforms which results in improved simultaneous range and velocity resolution, although their practical use case is limited \cite{Klauder_Bell_1960_01,Klauder_Bell_1960_02}. The goal in these early works on information theory for radars was information theoretic formulation in waveform design, to improve the performances in terms of target detection and parameter estimation \cite{Klauder_Bell_1960_01, Bell_TIT_1993, Yang_2007}. 
%Frost and Shanmugan computed the information content in the synthetic array radar images \cite{Frost_AES_1983}. 
The relationship between the mutual information between the target parameters of interest and the transmitted signal, and accuracy of parameter estimation was exploited in \cite{Bell_TIT_1993}, to design optimal radar waveforms. 
%Furthermore, parameterization of targets based on their respective target impulse responses, allows one to design such optimal transmit waveforms. 
Transmit waveform optimization for frequency diverse array (FDA) radar was considered in \cite{Nitsan_EUSIPCO_2019}, where target localization was done by analyzing the CRBs, and the knowledge of location of the targets was used to design the transmit waveform for FDA and to create range-angle dependent beams. Transmit array sub-aperturing was employed in FDA radar \cite{Gui_STSP_2018}, where each sub-array was assigned with different carrier frequencies and their weights were chosen adaptively via cognitive beamforming.

Even in the context of radars, the mean squared error (MSE) for parameter estimation is well-studied in literature. In \cite{Dogandzic_2001}, CRB for the target range, velocity, and angle-of-arrival (AoA) with a narrow band assumption on the transmit waveform was derived. Further, it was shown that the CRB for AoA is independent of the delay and velocity parameters. Moreover, CRB for AoA was shown to be a function of sensor locations only, through the moment-of-inertia parameters of the sensory array. It was noted in \cite{Rendas_1998} that the CRB and ambiguity functions were related to each other and impact the performance of parameter estimation, as follows -- ``\emph{The ambiguity function establishes global conditions under which the local bounds are accurate predictions of the expected error performance and identifies the regions of the parameter space where large errors may occur.}'' \cite{Rendas_1998}. Exploiting this key result, we extend the notion of ambiguity functions to wideband signals and general array geometries for our problem at hand. 

Bounds on data rate between a JRC receiver and a communication transceiver was studied in \cite{Chiriyath_SP_2016}, where the radar aids to relay the data from the communication transmitter to the receiver which is not co-located with the radar.\footnote{\textcolor{black}{Note that this setup is different from our scenario. In our setup, the radar transceiver is capable of sending and receiving data from the targets. The model in \cite{Chiriyath_SP_2016} assumes a different data transmission node, which is independent of radar and targets assist in relaying the information to the receiver which is collocated with radar receiver.}} This system follows a DFRC model, in which the radar performs joint data decoding and parameter estimation of targets using the combined received signal and then modulates the information onto the radar signal. Thus, the communication and radar receivers are able to decode the message and estimate the target parameters, respectively. Further, the data rate and \emph{parameter estimation rate} were balanced similar to the case of a two user multiple access channel at the radar receiver, and the corresponding rate region was derived. In \cite{Liu_TCom_2020}, a massive MIMO DFRC architecture was proposed, that employs hybrid beamforming to enable joint radar tracking and data communication. A radar transmission signal was designed using orthogonal frequency division multiplexing (OFDM) in \cite{Garmatyuk_Sensors_2011}. Here, uncertainties in the received data due to data symbols at the radar were mitigated, by dividing the target echo signal by apriori known or decoded data symbols in the frequency domain. In this setup, the transmitter and receiver are not co-located as in the case of bistatic radars. In \cite{Kumari_2020}, a joint waveform design using preambles of \textcolor{black}{the communication signal} as radar detection waveforms were employed to achieve a trade-off in the performance of data transmission and parameter estimation. Even though the usage of communication preambles as radar waveforms allows one to reuse the spectral and temporal resources, it results in non-optimal radar performance due to poor ambiguity function in the transmit waveform. For the preamble duration, multiple smaller periodic waveforms can be designed to achieve a better performance, especially when the targets move rapidly. \textcolor{black}{An experimental validation of data communication using BPSK modulation on radar with FMCW waveforms via backscattering by semi-passive targets, is reported in \cite{Cnaan-On_MTT_2015}, where single antenna was used at both ends. This results in a limited data rate for a given distance between the targets, due to the potential overlap of the spectrum between the transmission from two adjacent targets.}

\subsection{Contributions}  \label{SubSecContributions}
In this work, we derive lower bounds on the mutual information for the forward and reverse channels between a radar transceiver and targets, and derive the CRB on the minimum MSE of target parameters, namely range, velocity and AoA. We consider generic linear models for forward and reverse channels and target response function, for analytical tractability. The proposed model accommodates both narrowband and wideband transmit signals, any receiver antenna array geometry, and applicable for both monostatic and bistatic radars. For brevity, we consider monostatic radars in this paper.
%Towards this end, we derive analytical expressions for lower bounds on data rates and parameter estimates for the cases of (a) radar-only, (b) communication-only, and (c) joint radar and communication scenarios for the mono-static radar case. In particular, we compute the Fisher information matrix (FIM) to further calculate the CRB. 
Motivated by the relationship between CRB and the ambiguity function studied in \cite{Rendas_1998}, we propose novel target ambiguity functions (TAFs) for each target parameter, which can be optimized to achieve the corresponding CRB. We illustrate the utility of the proposed framework, which includes (a) optimal waveform design for JRC that minimizes TAFs, (b) JRC using overlaid data pulses on radar waveforms. \textcolor{black}{In summary,}
\begin{itemize}
    \item We derive the CRB on the minimum MSE of the estimates of relevant target parameters, for radar-only and JRC scenarios. \textcolor{black}{Three novel TAFs for range, velocity and azimuth -- which embed parameters such as elevation angle, target impulse responses and array geometry -- are introduced, which are important for the MSE bounds.}
    \item \textcolor{black}{Using calculus of variations, we propose an optimal waveform design procedure that minimizes a chosen TAF.} 
    %Via numerical example, Range TAF optimization is demonstrated}.
    \item We derive lower bounds on data rates over the forward and reverse channels between a JRC transceiver and a passive target.
    %which is shown to depend on the radar cross section (RCS) and the target response function.
    \textcolor{black}{Further, we show that communication on the forward channel does not affect the CRB. We also elaborate on the inter-target interference suppression and its effect on parameter estimation on reverse channel.}
    %However, data communication on the reverse channel can have an adverse effect on parameter estimation if the inter-target interference suppression is not sufficient. The connection between the TAFs and inter-target interference suppression is established.
    \item \textcolor{black}{Through numerical examples, we demonstrate the practical utility of the proposed methods in a JRC system in terms of transmit waveform optimization, beamformer design, Gaussian codebook design and data rate bounds.}
\end{itemize}

\subsection{Organization} \label{SubSecOrganization}
The remainder of the paper is organized as follows. In Section~\ref{sec:datamodel}, we discuss the data model and formulate the JRC problem with semi-passive targets. In Section~\ref{sec:EstBounds}, we compute the CRB for estimates of radar parameters for both radar-only, and JRC scenarios. In Section~\ref{sec:application}, we present the optimal transmit waveform design using calculus of variations. In Section~\ref{sec:dataBounds}, we derive lower bounds on communication data rate for forward and reverse channels. Section~\ref{sec:simResults} demonstrates the practical application of the proposed methods via numerical examples in JRC, receiver beamformer design, codebook design and transmit waveform optimization. Concluding remarks are provided in Section~\ref{sec:Conc}. 
%We provide numerical results on examples of the proposed framework in , where transmit waveforms are analysed through the target ambiguity functions. 

\begin{figure}[!t] 
  \begin{center}
  \includegraphics[width=3.35in]{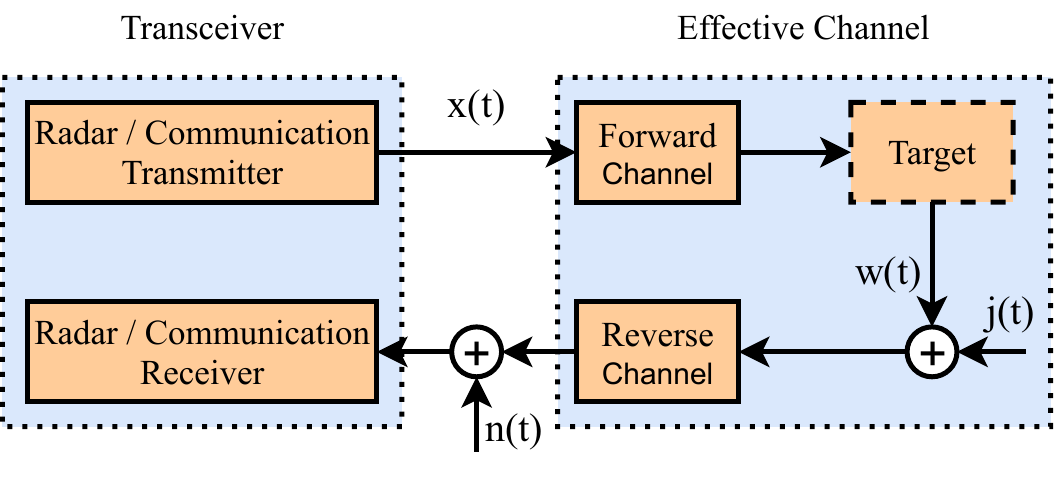}
 \end{center}
\caption{Block diagram of a joint radar and communication system with a transceiver and a single passive target.}
\label{fig:rit_model}
\end{figure}

\section{Data Model and Problem Formulation} \label{sec:datamodel}
 \subsection{Joint Radar and Communication Transmit Signal}  \label{SubSecRadarTx}
Consider a radar system as depicted in Fig.~\ref{fig:rit_model}, consisting of a  transceiver equipped with a single antenna joint radar and communication (JRC) transmitter, and a JRC receiver with $L$ antennas. The radar transmits a signal $x_i^{\text{RF}}(t)$ \textcolor{black}{at center frequency $f_c$}, for $i=1,\ldots,N$ with $N$ pulses of $\ts$ duration each. The average transmit power is $\pavg$, and the total transmit power $\ptx=N \ts \pavg$. The baseband component of $x_i^{\text{RF}}(t)$, denoted by $x_i(t)$, is assumed to be band-limited to $B$ Hz. In other words, magnitude spectrum of $x_i(t)$, denoted by $|X_i(f)|^2$, is positive for $|\omega| \leq B$, and zero otherwise. Throughout this paper, we consider our analysis on $x_i(t)$, even though all the channel effects such as Doppler shift, target impulse response, etc.~are applied on $x_i^{\text{RF}}(t)$. Further, we assume that the transmitted signal comprises of $N$ Gaussian pulses with $B\ts>1$ \cite{Woodard_RRE_1967}.\footnote{\textcolor{black}{We consider the Gaussian pulse for the ease of analysis, as opposed to other popular choices such as a rectangular pulse with a given duty cycle. Moreover, Gaussian modulated pulses are often used in practice \cite{Uduwawala_2007}, \cite{Rouhollah_MTT_2020}.} } That is,
\begin{equation}
x_i(t) = \frac{\sqrt{\pi}}{\beta} \exp\left(-\frac{\pi^2 (t-iT_s)^2}{\beta^2} \right), ~~~ \frac{-\ts}{2} \le t \le \frac{\ts}{2},
\label{eqn:gaussPulse}
\end{equation}
where the choice of $\beta$ determines the $3$ dB pulse bandwidth. The magnitude of the Gaussian pulse in frequency-domain representation can be written as  
\begin{equation}
|X_i(f)| = \exp\left(-\beta^2 f^2\right).
\end{equation}
Following \cite{Bell_TIT_1993}, we assume that the characteristics of $k^{\text{th}}$ target is captured in its random target channel response, $g_k(t), k=1,\ldots, K$, with the following assumptions.
\begin{enumerate}[(i)]
\item Finite energy: $g_k(t)$ satisfies  $\int_{-\infty}^{\infty} \mathbb{E} |g_k(t)|^2 \text{dt} < 1$, where $\mathbb{E}X$ denotes the expectation of a random variable $X$. Further, the response $g_k(t)$ does not include parameters such as the reflection coefficient and the effective radar cross section (RCS) of the $k^{\text{th}}$ target.
\item Causality: $g_k(t)$ is causal, i.e.~$g_k(t) = 0,~ t < 0$.
\item Fourier transform of $g_k(t)$, denoted by $G_k(f)$, exists.
\item Independence: $g_k(t)$ is independent across $k=1,\ldots,K$, and is independent of $x_i(t)$, for $i=1,\ldots,N$. 
\end{enumerate}

\textcolor{black}{In the considered setup, data communication along with radar target detection is achieved through} a random code $\textbf{c}$ of length $N$, \textcolor{black}{whose $i^{\text{th}}$ component is multiplied with $x_i(t)$} before transmission. \textcolor{black}{Vector \textbf{c}} is determined \textcolor{black}{by a sequence of Gaussian random variables} $\{a_i, i=1,\ldots N\}$.\footnote{\textcolor{black}{In radar-only mode, the pulses can be directly transmitted with maximum amplitude, without this modulation. In such cases, these unknown pulse amplitudes of are considered as nuisance parameters. Additionally, it is also possible that some fraction of the $N$ pulses can be retained for radar-only, while the remaining pulses can be modulated for JRC.}} Therefore, a total of $N B \ts$ resource dimensions are available to be shared between radar and communication. 
%\textcolor{blue}{If the transmitter uses all $N$ pulses,} 
\textcolor{black}{We assume that \textbf{c} is an $N$-dimensional Gaussian random vector} with zero mean vector and a covariance matrix $\sigma_x^2 \mathbf{I}_N$, where $\mathbf{I}_N$ denotes the identity matrix of size $N$, \textcolor{black}{and $\sigma_x^2=\pavg$}. \textcolor{black}{We assume that the radar transmitter and targets follow time division multiplexing for communication, where the transmitter sends data on odd frames and targets respond on even frames.} %Therefore, the information in each pulse is given by $\frac{1}{2} \log (2\pi e \sigma_x^2)$ nats.
%The end-to-end system model is described in Fig.~\ref{fig:target_channel_model}, based on which we next 
\textcolor{black}{Next, we} present the models for the reflected signal from $K$ \textcolor{black}{semi-}passive targets. 

\subsection{Reflected Signal from Semi-Passive Targets} \label{SubSecRxSignal}
\begin{figure*}[!t]  
  \begin{center}
  \includegraphics[width=7in, height=3cm]{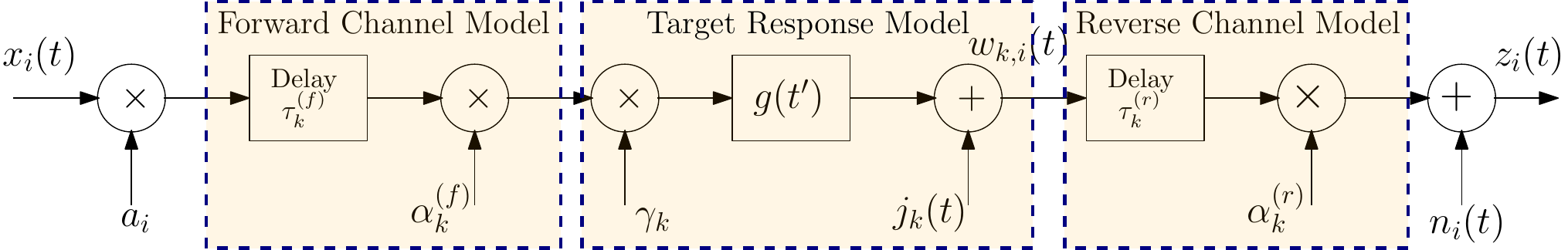}
 \end{center}
\caption{Linear target response model.}
\label{fig:target_channel_model}
\hrulefill
\end{figure*}

Figure~\ref{fig:target_channel_model} depicts the linear models of the forward and reverse channels between the radar transmitter and the $k^{\text{th}}$ target, and the corresponding target response channel. \textcolor{black}{In this work, we let the targets to be \emph{semi-passive}, i.e.~they can receive and modulate the data sent by radar transmitter -- using techniques such as ABC, or IRS-aided beamforming -- by using the active elements, but cannot transmit data using active components. Further, we assume that a strong direct line-of-sight path exists between the radar and targets. Therefore, the signal received via the multi-path reflections are relatively weak and can be ignored. However, our analyis can be extended to a general setup in a straightforward manner.} \textcolor{black}{Note that the random code $\mathbf{c}$ can be applied either at the transmitter or by the passive targets, or both.} First, let us consider a single antenna at the radar receiver, i.e.~$L=1$. The reflected signal from $k^{\text{th}}$ target for the $i^{\text{th}}$ pulse can be written as
\begin{align}
w_{k,i}(t) = a_i \alphakf \gamma_k \int g_k(t') x_i\left(t- t'-\taukf\right) dt',
\end{align}
where $\taukf$ is the propagation delay in the forward channel -- from the transmitter to the target, $\alphakf \triangleq \epsilon\ r_k^{-e}$ denotes the pathloss in the forward channel with $e$ as the pathloss exponent, and $\gamma_k$ denotes the reflection coefficient for $k^{\text{th}}$ target including the RCS of the target. For convenience, this continuous time system can be approximated by a discrete model as
\begin{align}
\int g_k(t') x_i\left(t- t'-\taukf\right) dt' \approx  \sum_{p=1}^D \zeta_{k,p} \delta\left(t-t_{k,p}\right), \label{eqn:DiscreteAppx}
\end{align}
with $D$ denoting the number of delay taps, and $\zeta_{k,p}$ being the amplitude in the $p^{\text{th}}$ tap due to $k^{\text{th}}$ target. \textcolor{black}{The delays between the multiple taps are negligible compared to the round trip delay between the transmitter and targets, and hence can be approximated as a single point reflection integrating all the energy from the taps. However, the extended targets which are in close proximity of the radar can be split into multiple targets moving together, without loss of generality. Optimal radar waveforms can be designed either in a multi-path environment or for detection of specific targets with known target response models, using this tap-delay model \cite{Bell_TIT_1993}.} The signal received at the radar due to non-moving targets can be written as 
\begin{align}
& z_i(t) = a_i \alphakr  \left( \sum_{k=1}^K \alphakf \gamma_k w_{k,i}(t - \taukr)  \right. \nonumber \\
& ~~~~~~~~~~~~~~ \left. + j_i(t-\taukr) \right) + n_i(t), ~~~ \frac{-\ts}{2}\le t \le \frac{\ts}{2},
\label{eqn:rx_eqn1}
\end{align}
where $\taukr$ and $\alphakr$ denote the propagation delay and the pathloss factor in the reverse channel, $n_i(t)$ denotes the additive white Gaussian noise (AWGN) with zero mean and unit variance, and $j_i(t)$ denotes the jamming signal. For the ease of analysis, we assume that the transmitter and receiver are collocated, in which case $\tau_k^f = \tau_k^r = \tau_k$ and $\alphakr = \alphakf = \alphak$. Therefore, \eqref{eqn:rx_eqn1} simplifies to
\begin{align}
& z_i(t) =  a_i \left( \sum_{k=1}^K \alpha_k^2 \gamma_k w_{k,i}(t \hspace{-0.1cm} - \hspace{-0.1cm} \tauk) \hspace{-0.1cm} + \hspace{-0.1cm} j_i(t \hspace{-0.1cm} - \hspace{-0.1cm} \tauk) \right) \hspace{-0.1cm} + \hspace{-0.1cm} n_i(t).
\label{eqn:rx_eqn2}
\end{align}
On the other hand, if the $k^{\text{th}}$ target is moving at a constant speed $v_k$ radially from the radar transmitter, we can write
\begin{align}
& z_i(t) =  a_i   \left( \sum_{k=1}^K \alphak^2 \gamma_k w_{k,i}(\left[t- \tauk\right] \mathbb{T}_k ) \right. \nonumber \\
& ~~~~~~~~~~ \left. + j_i(t-\tauk) \right)  + n_i(t), ~~~ \frac{-\ts}{2}\le t \le \frac{\ts}{2}, \label{eqn:rx_eqn_2}
\end{align}
where 
\begin{align}
\mathbb{T}_k = \sqrt{\frac{c+2 v_k}{c-2 v_k}} \approx \frac{c + v_k}{c - v_k},
\end{align}
denotes the time-axis compression factor due the Doppler shift in frequency from a moving target.\footnote{\textcolor{black}{Note that this is a generic model which includes both narrow-band and wide-band transmitted signals. Moreover, for $L>1$, this model is independent of the array geometry. It is easy to show that for $v_k \ll c$, the difference between the observed frequency $f_o$ and $f_c$ simplifies to $f_c - f_o =(1-\frac{1}{\mathbb{T}_k})f_c \approx \frac{2~v_k f_c}{c}$.}} Now, separating the reflected signal from  $k^{\text{th}}$ target, we get \eqref{eqn:rx_eqn3}, which is given in the top of next page.
\begin{figure*}
\begin{align}
z_i(t) =  a_i \alphak^2 \gamma_k w_{k,i}([t- \tauk]\mathbb{T}_k) + a_i \underbrace{\left( \sum_{j \neq k} \alpha_j^2\ \gamma_j\ w_{j,i}([t- \tauj]\mathbb{T}_i) + j_i(t-\tauj) \right)}_{\texttt{Interference for $k^{\text{th}}$ target}} + n_i(t). \label{eqn:rx_eqn3}
\end{align}
\hrulefill
\end{figure*}
Note that the reflection from each target may undergo different time compression, depending on the velocity of the target. Moreover, as discussed earlier, it is possible that some of the transmitted data pulses from these passive targets can be used for sending data information, while the remaining can be used to detect the presence of a target and estimate its parameters. The trade-off between the performance of radar target parameter estimation and the data transmission is viable by proportioning the available resources such as energy, bandwidth and codes. In this sense, it is not necessary that all the $N$ transmitted pulses need to be identical. However, set of all possible realizations of $x_i(t)$ comes from the finite \textcolor{black}{Cartesian product} of waveforms selected from two ensembles; one corresponding to radar parameter estimation and another for data transmission.

Next, we consider the general case where the receiver has $L>1$ antennas, and the received signal vector is written as
\begin{align}
\mathbf{z}_i(t) = a_i \mathbf{X}_i(t) \mathbf{\Gamma}_i(t) + \mathbf{n}_i(t),  \label{eqn:zrxvec}
\end{align}
where 
%$a_i$ is a random variable representing amplitude of the $i^{\text{th}}$ Gaussian pulse, 
$\mathbf{X}_i(t)$ is a $L \times K$ matrix, whose $(l,k)^{\text{th}}$ entry is
\begin{align}
    \mathbf{X}_i^{\text{\footnotesize (l,k)}}(t)=x([t- i \ts -2 \tauk -\phil]\mathbb{T}_k),
\end{align}
for $k=1,\ldots,K$, $l=1,\ldots,L$, where $\varphi_l \triangleq \frac{\mathbf{p}_k^T \mathbf{p}_l}{c}$, with $\mathbf{p}_l$ denoting the relative Cartesian coordinates --- with respect to a reference element in the receive antenna array as origin --- of the $l^{\text{th}}$ antenna element and $\mathbf{p}_k$ are the direction cosines of the $k^{\text{th}}$ target with respect to the antenna. Using \eqref{eqn:DiscreteAppx}, the $k^{\text{th}}$ entry of the $K$-length vector $\mathbf{\Gamma}_i(t)$ can be written as 
\begin{align}
    \mathbf{\Gamma}_i^{\text{\footnotesize (k)}}(t) = \gamma_k \alphak^2 \sum_{p=1}^D \zeta_{k,p} \delta\left(t-t_{k,p}\right).
\end{align}
The $L$-length vector $\mathbf{n}_i(t)$ is a spatio-temporally white Gaussian random vector with mean being the zero vector and covariance matrix $\sigma_n^2 \mathbf{I}_L$. The mean vector of $\mathbf{z}_i(t)$ for all $i=1,\ldots,N$ is also assumed to be the zero vector. \textcolor{black}{In the next section, we derive lower bounds on parameter estimation for range, velocity and angle in the forward and reverse channels.}

\section{Bounds on Parameter Estimation} \label{sec:EstBounds}
\subsection{Radar-Only Transmission} \label{sec:RadarEstBounds}
The Cram\'er-Rao lower bounds (CRB) on the variance of the radar target parameters in our setup can be obtained by finding the inverse of the Fisher information matrix (FIM). Following \eqref{eqn:zrxvec}, recall that the $L$-length vector $\mathbf{z}_i$ follows a Gaussian distribution with mean vector $a_i \mathbf{X}_i \mathbf{\Gamma}_i$, and covariance matrix $\mathbf{R}_n \triangleq \mathbb{E}\left(\mathbf{n}_i \mathbf{n}_i^T\right)$, independent across $i=1,\ldots,N$. Let $\mathbf{\Theta} \triangleq [\mathbf{r}~~\mathbf{v}~~\boldsymbol{\phi}]^T$ denote the vector of desired parameters, and $r_k$, $v_k$ and $\phi_k$ denote the $k^{\text{th}}$ entries of vectors $\mathbf{r}$, $\mathbf{v}$ and $\boldsymbol{\phi}$, respectively, which correspond to the range, Doppler and azimuth angle of the $k^{\text{th}}$ target.\footnote{\textcolor{black}{In a radar-only scenario, nuisance parameters such as $\theta$, $\mathbf{c}$, i.e., $\{a_i\}$, tap coefficients and amplitude of $g_k(t)$ exist. Tighter CRBs can be computed when these are conditioned out in the computation of FIM, which will not be discussed in this paper.}}

Towards this end, the second order partial derivatives of the logarithm of joint PDFs of $\textbf{z}_i, i=1,\ldots,N$ computed with respect to the parameters $r_k$, $v_k$ and $\phi_k$, $k=1,\ldots,K$ are derived in Appendix~\ref{app:taf}, and the final expressions are given in equations \eqref{eqn:pd_range0}-\eqref{eqn:pd_angle0} at the top of this page.
\begin{figure*}
\begin{align}
& \mathbb{E}\left[\frac{\partial^2 \log f_z(\mathbf{z}_i)}{\partial r_k^2}\right]   = - \frac{16 \rhot N \pi^4 \mathbb{T}_k^4 ~ \mathbb{E}[\gamma_k^2] ~ \mathbb{E}[\|\zeta_k\|^2] \epsilon^4 r_k^{-4e}}{\beta^4\ c^2} \mathbb{E} \left( \|  \mathbf{x}_{i,k}  \circ \mathbf{t}_k \|^2 \right) .  \label{eqn:pd_range0} \\
& \mathbb{E}\left[\frac{\partial^2 \log f_z(\mathbf{z}_i)}{\partial v_k^2}\right]   = - \frac{16  \rhot N \pi^4 c^2 \mathbb{T}_k^2 ~ \mathbb{E}[\gamma_k^2] ~ \mathbb{E}[\|\zeta_k\|^2] \epsilon^4 r_k^{-4e}}{\beta^4\ (c-v_k)^2} \mathbb{E} \left( \|  \mathbf{x}_{i,k}  \circ \mathbf{t}_k \circ \mathbf{t}_k \|^2 \right). \label{eqn:pd_velocity0} \\
& \mathbb{E}\left[\frac{\partial^2 \log f_z(\mathbf{z}_i)}{\partial \phi_k^2}\right]   = - \frac{4 \rhot N \pi^2 \mathbb{T}_k^4 ~ \mathbb{E}[\gamma_k^2] ~ \mathbb{E}[\|\zeta_k\|^2]\epsilon^4 r_k^{-4e} }{\beta^4} \mathbb{E} \left(  \| \mathbf{x}_{i,k}  \circ \mathbf{t}_k \circ \mathbf{\Phi}_k \|^2 \right).  
\label{eqn:pd_angle0}
\end{align}
\hrulefill
\end{figure*}
Here, $\circ$ denotes the Hadamard (or element-wise) product of two matrices of same size, $\rhot={\left(\frac{\mathbb{E}[a_i^2]}{\sigma_n^2}\right)}$ is the transmit SNR, $c$ is the speed of light, $\mathbf{x}_{i,k}$ is the $k^{\text{th}}$ column of $\mathbf{X}_i$, and $\mathbf{\Phi}_k$ is defined as
\begin{align}
\mathbf{\Phi}_k \triangleq \frac{\left\{ \sin \theta_k \left(y_l \cos \phi_k - x_l \sin \phi_k \right) + z_l \cos \theta_k \right\}}{c},
\end{align}
for $l=1,\ldots,L$ is a vector of dimension $L \times 1$. The Cartesian coordinates of the $l^{\text{th}}$ receive antenna is denoted by $(x_l, y_l, z_l)$ and $\mathbf{t}_k$ is the vector of relative time instants at which the wavefront from $k^{\text{th}}$ target hits the receiver antenna elements. 

It can be observed that the lower bound on  the MSE across all the parameters depend on the key factors such as $N$, $\rhot$, RCS value $\mathbb{E}[\gamma_k^2]$, extension nature of the target $\mathbb{E}[\|\zeta_k\|^2]$, and the target ambiguity functions (TAF). Therefore, by designing a radar transit waveform which minimizes the TAFs, lower bounds on the minimum MSE can be written as
\begin{align}
& \mathbb{E}[(r_k - \hat{r}_k)^2] \ge \frac{-1}{\mathbb{E}\left[\frac{\partial^2 \log f_z(\mathbf{z}_i)}{\partial r_k^2}\right]}, \\
& \mathbb{E}[(v_k - \hat{v}_k)^2] \ge \frac{-1}{\mathbb{E}\left[\frac{\partial^2 \log f_z(\mathbf{z}_i)}{\partial v_k^2}\right]}, \\
& \mathbb{E}[(\phi_k - \hat{\phi}_k)^2] \ge \frac{-1}{\mathbb{E}\left[\frac{\partial^2 \log f_z(\mathbf{z}_i)}{\partial \phi_k^2}\right]} .
\end{align}
Note that the cross terms in the FIM are ignored to derive the lower bounds in \eqref{eqn:pd_range0}-\eqref{eqn:pd_angle0}. In general, the achievable CRBs (local bounds) are influenced by the ambiguity functions (global conditions), as observed in \cite{Rendas_1998}. That is, if one designs waveforms that achieve the least possible ambiguity functions, the corresponding CRBs listed in \eqref{eqn:pd_range0}-\eqref{eqn:pd_angle0} becomes achievable. Therefore, the approach considered here -- optimizing the ambiguity functions -- is flexible, in the sense that the performance of the radar can be designed by individual optimization of the TAF for range, velocity and angle, for selected set of targets within the given range, velocity or angle-of-arrivals. This is advantageous in contrast to the conventional ambiguity functions defined to optimize all parameters using one transmit waveform function. \textcolor{black}{As as example, the process the optimizing the range TAF is detailed later in Sec.~\ref{subsec:txwaveform}. Other TAFs can be optimized along similar lines.} \textcolor{black}{See Appendix~\ref{app:taf} for a detailed discussion on TAFs.}

\subsubsection{MSE on Range Estimation}
Note that the CRB for range estimation given in \eqref{eqn:pd_range0} decreases with $\rhot$, $N$, mean RCS value $\mathbb{E}[\gamma_k^2]$, mean energy in the target impulse response $\mathbf{\zeta}_k$ as well as the effective BW of the signal $\int_{t=-\frac{T_s}{2}}^{\frac{T_s}{2}} \left(\frac{d x(t)}{dt}\right)^2 dt$, as observed in \cite{SMKay_Book_VolI_1993}. Another interesting observation is that the bound increases significantly with large velocity values, i.e.~when $\mathbb{T}_k < 1$, but only increases as $1/ \mathbb{T}_k^4$ for low velocities. The parameter $\beta$ is inversely proportional to $B$, and hence the bound decreases with fourth power on $B$, which is important to note. 

\subsubsection{MSE on Velocity Estimation}
From \eqref{eqn:pd_velocity0}, it can be observed that the velocity estimation also improves with the previously mentioned parameters. The key difference is observed in the new term $\int_{t=-\frac{T_s}{2}}^{\frac{T_s}{2}} \left(\frac{d^2 x(t)}{d\ t^2}\right)^2 dt$ in place of the effective bandwidth term in  \eqref{eqn:pd_range0}. With large bandwidth, this quantity will increase exponentially and improves the velocity estimation, provided the wide-band array configuration is used as in the case of space-time adaptive processing (STAP) filters \cite{Ward_ICASSP_1995}. That is, using narrow-band pulses, or shortening the array aperture to meet the narrow-band conditions greatly impact the velocity and angle estimation. 

\subsubsection{MSE on Angle Estimation}
Apart from the effect from the expected parameters, the key difference in this case is the influence of $\mathbf{\Phi}_k$ due to array geometry, as seen in \eqref{eqn:pd_angle0}. 
%This influence can be understood by studying the angle TAF defined in Appendix~\ref{app:taf}. Also, Note that the absence of array geometry parameters in the range and velocity MSE bounds directly, but it appears indirectly in the range TAF and velocity TAF functions. 
\textcolor{black}{Further, it is important to note that it may appear that the array geometry may not influence the bounds in \eqref{eqn:pd_range0} and \eqref{eqn:pd_velocity0}. This is not true, as the TAFs corresponding to range and velocity do depend on the array geometry, as discussed in Appendix~\ref{app:taf}. This is an interesting result emerging from our analysis, in contrast to the existing results, e.g.~\cite{Dogandzic_2001}. In most of the analysis in the existing literature, the array geometry does not influence the CRBs for range and velocity estimation, and the CRB on the angle estimation error does not depend on the range and velocity estimation errors.}

\subsection{Joint Radar and Communication}
%It can be observed from rate equations (\ref{eqn:forwardRate}) and (\ref{eqn:reverseRate}), and limits on MSE  in 
\textcolor{black}{It can be observed from the equations}
\eqref{eqn:pd_range0}-\eqref{eqn:pd_angle0} \textcolor{black}{that the bounds on MSE are linear in $N$. The corresponding CRBs on MSE on the forward and reverse channels of an JRC system can be computed as follows.}

\subsubsection{JRC on the Forward Channel}
Consider the scenario where radar pulses are modulated by the $N$-dimensional Gaussian code $\textbf{c}$ designed in Appendix~\ref{app:codeConstr}. At the radar receiver, the amplitudes $\{a_i, i=1,\ldots,N\}$ are known and hence can be used in the parameter estimation. \textcolor{black}{Therefore, the previously computed parameter estimation bounds given in \eqref{eqn:pd_range0}-\eqref{eqn:pd_angle0} are applicable even in this case. This result is in contrast with the work in \cite{Chiriyath_SP_2016}, where the JRC transmitter and receiver are not co-located, i.e.~bi-static radar. However, our results can also be extended to the scenario in \cite{Chiriyath_SP_2016} as follows. The parameter estimation at the JRC receiver, and information transmitted to the communication receiver using data pulses overlaid with the radar pulses can be treated as part of a two-user multiple access channel, and the CRBs can be computed \cite{TomCover_Book_1991}.}

\subsubsection{JRC on the Reverse Channel}
In this case, radar transmits constant modulus pulses \textcolor{black}{during the frame in which the targets are expected to send data}, and the targets modulate them using their backscattering control elements. Let $\{ b_{ki}, i=1, \ldots, N$ \} denote the reflection amplitudes due to data modulation from the $k^{\text{th}}$ target to the radar receiver. 
%The radar-data receiver compares the amplitude of the pulses at the peaks of the received signal and uses them to compare with the scaled codebook made known to it, apriori. The Euclidean distance is computed for minimum distance decoding. The radar receiver then compensates for the amplitude variation in the received pulses as per the decoded data. Thus, amplitude corrected data can be used to estimate the target parameters. 
%\textcolor{black}{
%The radar first estimates the target parameters from the received pulses after ignoring the modulation of the pulses. As evident in the sequel, incoherent averaging over several pulses can remove the influence of the code, on the parameter estimation provided there is separation between the targets in range, velocity and angle of arrival. When the targets are close to each other, the CRB have to be recomputed with data bits as additional unknown parameter in the FIM. If the decoded data bits are used to refine the parameter estimation as done in some of the literature \cite{}, there will be adverse effect when there are bit errors.  Once the target parameters are estimated, they are used provide filtering available in each of the parameter domains namely range, doppler and angle. For details refer to Section~\ref{sec:simResults}. This is possible due to the fact that average energy in the pulses are kept same even after coding is done.}
To obtain the parameter estimation bounds for this case, we need to compute the partial derivatives with respect to symbols $b_{ki}$ additionally, and use them along with the derivatives for the radar parameters to invert the larger FIM. The second derivative with respect to the symbols $b_{ki}$ can be shown as
\begin{align}
\mathbb{E}\left[\frac{\partial^2 \log f_z(\mathbf{z}_i)}{\partial b_{ki}^2}\right] = -\rho_T \mathbb{E}[b_{ki}^2]  \mathbb{E}[\gamma_k^2] ~ \mathbb{E}[\|\zeta_k\|^2] \nonumber \\
\epsilon^4 r_k^{-4e}  \mathbb{E} [ \| \mathbf{x}_{i,k} \|^2 ],
\end{align}
and the cross-term second order derivative can be written as
\begin{align}
\mathbb{E}\left[\frac{\partial^2 \log f_z(\mathbf{z}_i)}{\partial b_{ki} \partial r_k}\right] = -\rho_T 4 \pi^2 \mathbb{E}[b_{ki}^2] \mathbb{T}_k^2 \mathbb{E}[\gamma_k^2] ~ \mathbb{E}[\|\zeta_k\|^2] \nonumber \\
\epsilon^4 r_k^{-4e}  \mathbb{E} [ (\mathbf{x}_{i,k} \circ \mathbf{t}_{ik})^T \mathbf{x}_{i,k}],
\end{align}
for one parameter. \textcolor{black}{The derivatives in a general case can be derived similarly. It can be noticed that the cross-terms are not zero and hence impact the inverse of FIM and the MSE bounds. That is, the decision on each and every bit transmitted by the individual targets impact the performance all radar parameters for all targets. Even for a single target, the FIM for all parameters and data does not simplify and all data decision errors impact the performance of the radar. Hence, some level of orthogonality in resource allocation is a must for getting good performance in terms of bit error rate (BER) and parameter estimation error, where previously computed data rate bounds and parameter estimation bounds remain valid. }

\section{JRC Transmitter and Receiver Design} \label{sec:application}
%using Calculus of Variations}  \label{subsec:txwaveform}
\textcolor{black}{Following the analysis in Sec.~\ref{sec:EstBounds}, one needs to minimize the TAFs to achieve the CRB.} Minimization of TAFs is equivalent to maximization of the average norm $\mathbb{E} [\| \mathbf{x}_k \circ \mathbf{t}_k \|^2]$, and minimization of the following three inner products:
\begin{enumerate}[(a)]
    \item $\mathbb{E}[(\mathbf{x}_{i,m} \circ \mathbf{t}_m)^T \mathbf{R}_{n}^{-1} (\mathbf{x}_{i,n} \circ \mathbf{t}_n)]$,
    \item $\mathbb{E}[(\mathbf{x}_{i,m} \circ \mathbf{t}_m \circ \mathbf{t}_m)^T \mathbf{R}_{n}^{-1} (\mathbf{x}_{i,n} \circ \mathbf{t}_n \circ \mathbf{t}_n)]$, and
    \item $\mathbb{E}[(\mathbf{x}_{i,m} \circ \mathbf{t}_m \circ \mathbf{\Phi}_m)^T \mathbf{R}_{n}^{-1} (\mathbf{x}_{i,n} \circ \mathbf{t}_n) \circ \mathbf{\Phi}_m]$,
\end{enumerate}
simultaneously, for $m, n=1,\ldots,K$ and $n \neq m$, where the expectation is taken over all possible values of ranges, velocities, and azimuth angles.\footnote{Recall that the conditions (a)-(c) follow our analysis with Gaussian pulses. \textcolor{black}{However, they are also valid for all continuously differentiable functions with finite support in time and frequency, e.g.~$n^{\text{th}}$ order spline, half-sine, and raised cosine pulses. For more details, refer to Section~\ref{sec:simResults}.}} 
%For other envelope waveform, the optimization criteria can be different. In this section, we substantiate the selection of Gaussian pulses, by analyzing the three TAFs.} 

\textcolor{black}{For $L \gg 1$, the above problem can be considered as a functional optimization, carried out through methods in calculus of variations \cite{CoV_SaganBook_1969}. Minimizing the inner product between time derivatives of the pulses corresponding to two different targets can be formulated as an optimization problem, as discussed in the next subsection. For the ease of presentation, we first use a continuous-time waveform, and relax that requirement later.}

\subsection{Optimization of Transmit Waveform} \label{subsec:txwaveform}
\textcolor{black}{Consider a function $f(t,y(t),\dot{y}(t))$ defined over an interval $a \le t \le b$, where $\dot{y}(t) \triangleq \frac{dy(t)}{dt}$. Succinctly represented as $f(t,y,\dot{y})$ for convenience, let the second order partial derivatives with respect to $t,y, \text{and}, \dot{y}$ be continuous. Then, the functional defined as 
\begin{equation}
F(y, \dot{y}) = \int_a^b f(t,y,\dot{y}) dt 
\end{equation}
}
\textcolor{black}{attains its extrema when
\begin{equation}    
\frac{\partial F}{\partial y} = \frac{d}{dt} \left( \frac{\partial F}{\partial \dot{y}}   \right) .
\label{eqn:EulerEqn}
\end{equation}
}
\textcolor{black}{In the context of TAFs, we define the following cost function $F$, as the normalized inner product}  \textcolor{black}{
\begin{equation}
F(y, \dot{y}) \triangleq \frac{1}{E}\int_{-\frac{T_s}{2}}^{\frac{T_s}{2}} \dot{y}(t) \dot{y}(t+\delta) dt, ~~ \delta > 0, 
\label{eqn:CostFunc}
\end{equation}}
\textcolor{black}{where $E \triangleq \int_{-\frac{T_s}{2}}^{\frac{T_s}{2}} \dot{y}(t) \dot{y}(t) dt$.}
\textcolor{black}{This is equivalent to evaluating the normalized auto-correlation of the first derivative of $y(t)$. Consider an iterative algorithm that finds an extremum of this function $F$. The smallest $\delta$ that minimizes $F$ such that $F^*(\delta) \le \epsilon$, for some $\epsilon > 0$, is considered as the smallest possible ambiguity value for that $\epsilon$. That is, the derivative of $y(t)$ should be similar to a Kronecker delta impulse function, such that $F(y,\dot{y}) < \epsilon$ for an arbitrary $\delta > 0$. Without loss of generality, the scale factor $1/E$ can be ignored during the optimization stage.}

%\begin{figure}[!t]  
%  \begin{center}
%  \vspace{-1.5in}
%  \includegraphics[width=4in]{threeEnvelopes.pdf}
% \end{center}
%  \vspace{-1.5in}
% \caption{Typical envelope waveforms used as starting points for optimization.}
%\label{fig:txWaveforms}
%\end{figure}

\textcolor{black}{For illustration, consider a simple case where $y(t)$ is one of the continuously differentiable baseband waveforms given in Table~\ref{tab:waveforms}, to start the optimization procedure. Here, $y(t)$ is the transmit pulse, which modulates a carrier signal such as frequency modulated continuous wave (FMCW) or a single-tone carrier, along with the amplitude scaling as per the code used. 
%These waveforms also illustrated in Figure~\ref{fig:txWaveforms}, in which all of them are normalized to have unit energy and have common support between $-\frac{T_s}{2}$ and $\frac{T_s}{2}$ with $T_s = 100 \mu sec$. 
Note that these envelope waveforms have a time support of $T_s$ seconds and frequency support of $B$ Hz. Let $y(t)$ evaluated at the $n^{\text{th}}$ step of the algorithm be denoted as $ y^{n}(t)$, and the algorithm stops when
%is perturbed by another continuously differentialble waveform such that the distance between the waveforms have small norm $\epsilon > 0$. That is,
$\| y^{n+1}(t) - y^{n}(t) \|_2 < \epsilon$, where $\|\cdot\|$ represents the Euclidean norm in the $\mathcal{L}_2$ space (or the Paley-Wiener space) for finite energy functions.
%is used as the distance measure between the two functions. 
Then, the following theorem specifies the condition under which the cost function in \eqref{eqn:CostFunc} converges. To summarize, if the perturbation on $y^n(t)$ is assumed to be a time-scaled, time-shifted and amplitude-scaled version of $y(t)$, such that the other parameters meet the above mentioned conditions, then the algorithm converges to an optimal waveform which minimizes the cost in \eqref{eqn:CostFunc}.} 
%\textcolor{black}{
%\begin{theorem}
%If $|y^{n+1}(t) - y^{n}(t)| \le \lambda~ y(\alpha t + \beta)$, then the condition for the convergence of the iterative algorithm is given by
%\begin{eqnarray}
%\left(\frac{\dot{y}_t}{y_t} + %\frac{\ddot{y}_t}{\dot{y}_t}\right)^2 + 4 \left( %\frac{\ddot{y}_t}{y_t}\right) \left[1 - \Gamma \right] \ge 0 
%\end{eqnarray}
%where
%\begin{equation}
%\Gamma \triangleq \frac{\left(1 %-\frac{\alpha^2}{\dot{y}_{t+\delta}} \right)}{\left(1  -  %\frac{\alpha^2}{\dot{y}_t} \right)} ,
%\end{equation}
%$\lambda > 0$, $\alpha > 0$ and $\beta > 0$ are design parameters which determining the speed of convergence. 
%\end{theorem}}
\textcolor{black}{
\begin{theorem}
Let $\|y^{n+1}(t) - y^{n}(t)\| \le \lambda y(\alpha t + \beta)$. Then, the condition for the convergence of an iterative algorithm is
\begin{eqnarray}
\left(\frac{\dot{y}_t}{y_t} + \frac{\ddot{y}_t}{\dot{y}_t}\right)^2 + 4 \left( \frac{\ddot{y}_t}{y_t}\right) \left[1 - \frac{\left(1 -\frac{\alpha^2}{\dot{y}_{t+\delta}} \right)}{\left(1  -  \frac{\alpha^2}{\dot{y}_t} \right)} \right] \ge 0,
\end{eqnarray}
and choices of $\lambda > 0$, $\alpha > 0$, $\beta \in \mathbb{R}, \beta \neq 0$, and $\delta > 0$ determine the speed of convergence.
\end{theorem}}
\begin{proof}
See Appendix \ref{app:cov}.
\end{proof}

\begin{center}
\begin{table*}
    \centering
    \begin{tabular}{|c||c|c|c|}
    \hline 
        Sl. & Envelope Type & Energy-Normalized $y(t)$   & Scaling Factors \\
        \hline \hline
        1   & Gaussian & $A  \exp\left(-\left[\frac{\pi~t}{\beta}\right]^2\right)$ & $A = \left(\frac{2 \beta^2}{\pi}\right)^{\frac{1}{4}} \sqrt{\text{erf}\left(\frac{\pi T_S}{\beta \sqrt{2} } \right)}$, $-\frac{T_s}{2} \le t \le \frac{T_s}{2}$, e.g.~$\beta = \frac{3}{4} T_S$ \\ 
            \hline 
        2 & Cubic spline & $A \left\{\begin{array}{ll} \frac{2}{3}-|t'|^2 + \frac{|t'|^3}{2} &, 0 \le |t'| < 1 \\
\frac{(2-|t'|)^3}{6} &, 1 \le |t'| < 2 \\
0 &, |t'| \ge 2 \end{array}  \right.$     & $t' = \frac{4~t}{T_s}$, ~~~$A = \sqrt{\frac{4}{T_s}}$\\ \hline
        3 & Half-sine wave & $A \sin\left(\frac{\pi t}{T_s}  \right)$ & $A = \sqrt{\frac{2}{T_s}}$ \\
        \hline        
    \end{tabular}
        \caption{Some well-known continuously differentiable transmit pulse envelopes.}    \label{tab:waveforms}
\end{table*}
\end{center}
\textbf{Discussion}: 
\textcolor{black}{Recall that the actual transmitted pulse will be a product of the envelope signal and a carrier. Towards waveform optimization, the product signal will only help improve the correlation properties of the waveform further. This will be detailed in Sec.~\ref{sec:simResults}, where the correlation properties between an FMCW-modulated optimized waveform and that of only the optimized waveform are compared. Further, recall that the TAFs are functions of the waveform samples, where the inner product between the samples at two different instants -- dictated by the array geometry -- is used as the cost function. As discussed in Appendix~\ref{app:arrayCorr}, the sampling time-instants vary with the angle of arrival and the array geometry. The derivative of the transmit waveform pulse will be sampled according to the angle of arrival for each target and the array geometry.\footnote{Here, the transmit and receive waveforms are assumed to be same for convenience. However, the waveforms will be different in a multi-path scenario, and the correlation of the corresponding derivatives have to be evaluated.} Having a large number of antennas yields a sufficient of number of samples to represent the entire waveform. However, for short arrays, one needs to optimize the correlation for those specific time instants averaged over all possible angles of arrival. The sampling instants of the received waveform is illustrated for two array geometries in Appendix~\ref{app:arrayCorr}. Such a discrete functional variation optimization is similar to the one presented above. Section~\ref{sec:simResults} details this procedure via numerical examples. 
%Moreover, Appendix~\ref{app:arrayCorr} demonstrates the correlation between the array vectors for two different array geomtries namely, uniform linear array (ULA) and uniform circular array (UCA). It also describes the nature of sampling time instants for the two array geometries. 
The discrete CoV can be performed for a given array geometry and the transmit waveform can be optimized by averaging the cost over all possible angles of arrival. Note that a waveform can also be optimized even if one restricts the array field of view (FoV) to a limited set of angles or sectors.
}

\textcolor{black}{
\subsection{Inter-Target Interference Suppression}
From the waveform optimization procedure described in the previous subsection, it is clear that the all the three TAFs can be reduced when the ranges, velocities and angles of the targets do not overlap. To quantify the impact of these reductions on the CRB of parameters, first consider the range estimation. In the case of a pulsed radar, the received waveform is correlated with the transmitted pulse and the time-lag at which correlation peak occurs is considered as the round-trip delay of that target. Therefore, any improvement in the auto-correlation of the transmitted waveform, will directly impact the energy leakage of one target onto the another. Even in the case of an FMCW radar, the transmitted FMCW waveform is mixed with the received waveform to obtain the tone frequency corresponding to the range of the target. The beat frequency in the mixer output is proportional to the delay in the reflected signal due to the nature of the FMCW signal. However, this is true only if the signal is not modulated. When the FMCW signal is modulated by another envelope signal, the correlation between the baseband equivalents of the transmitted and received pulse is calculated to detect the delays. This gives a better suppression of interference leaking from one target to another. In particular, the low pass filter after the mixing only gives a \emph{$\sinc(\cdot)$ roll-off} in comparison with the lower side-lobes obtained from a good auto-correlation in the transmitted waveform.}

\textcolor{black}{
Let $\sbr$ denote the side-lobe level obtained by the above transmit waveform optimization in the range dimension. That is, two objects separated by a small distance -- equivalent to round trip delay of $1/2B$ -- is recognizable by an FMCW signal with bandwidth $B$, will be suppressed by $\sbr$. In other words, we achieve a filtering of $\sbr$ across all ranges, as long as the two objects are not in close proximity. Next, even while processing the Doppler domain signal, a suppression is observed due to the $\sinc(\cdot)$ roll-off, even with a simple FFT processing. This suppression can be improved further by using, e.g.~a Hamming or a Blackman-Harris window. Let $\sbd$ denote this amount of suppression in the Doppler dimension.
Later, even with a simple FFT processing for angle estimation, a suppression is observed due to the array vector correlation between two angles. Let $\sba$ denote this suppression due to the array vector correlation. This array vector correlation is described in detail for two separate array geometries in Appendix~\ref{app:arrayCorr}, with no array tapering weights. Next, we will quantify the impact of these three suppression terms on the parameter estimation.
}

\textcolor{black}{
If two targets are closer in range, but not in Doppler and angle, then they can be separated out due to the filtering gain due to $\sbd$ and $\sba$. That is, even if the two targets have overlapping spectrum in the range FFT, a total interference suppression of $\sbd + \sba$ is achievable. For notational convenience, we represent the gain from these three filters as the product of three $\dsinc(\cdot)$ terms, i.e.~Dirichlet kernels. That is, the inter-target interference is reduced by, 
}
\textcolor{black}{
\begin{align}
& \sbtotal \approx  \dsinc(\Delta r) \, \dsinc(\Delta v) \, \dsinc(\Delta \theta), ~~~\text{where} \nonumber \\
& \dsinc(\Delta r) \triangleq \frac{1}{\nrange} \left\{\frac{\sin\left(\pi \Delta n  \nrange \right)}{\sin\left(\pi \Delta n \right)} \right\}, \nonumber \\
& \dsinc(\Delta v) \triangleq \frac{1}{\ndopp} \left\{\frac{\sin\left(\pi \Delta k \ndopp \right)}{\sin\left(\pi \Delta k \right)} \right\}, \nonumber \\
& \Delta n \hspace{-0.1cm} = \hspace{-0.1cm} \left \lfloor \frac{2 \Delta r B \nrange}{c F_r T_s} \right\rfloor, \Delta k \hspace{-0.1cm} = \hspace{-0.1cm} \left \lfloor \frac{\Delta f \ndopp}{F_d} \right \rfloor, \Delta f \hspace{-0.1cm} = \hspace{-0.1cm} \frac{2 \Delta v f_c}{c}, \nonumber \\
& \dsinc(\Delta \theta) \triangleq \frac{1}{L} \mathbf{a}^H(0) \mathbf{a}(\Delta \theta),
\label{eqn:sb_total}
\end{align}
where $\nrange$ and $\ndopp$ denote the size of the FFTs in the range and Doppler domains, respectively, $F_r$ and $F_d$ denote the sampling rates in the range and Doppler domains, respectively, and $\textbf{a}(\cdot)$ denotes the array steering vector. We use $\sbtot$ to denote the suppression between the $k^{\text{th}}$ and $j^{\text{th}}$ targets, with $\Delta^{(k,j)}_r \triangleq |r_k-r_j|$,
$\Delta^{(k,j)}_v \triangleq |v_k-v_j|$ and $\Delta^{(k,j)}_\theta = |\theta_k-\theta_j|$.
}

\subsection{Receiver Beamformer Design} \label{sec:beamWeights}
\textcolor{black}{
Given that it is possible to design a filter to separate two closely placed targets in the Doppler and angle domains even if their reflected signals have overlapping spectrum in the range domain, it is desirable to first design a digital beamformer at the receiver. Since the angles of targets are not known apriori, the FoV can be split into several sectors (beams) and weights for each beam to get an SINR improvement 
%both from other targets as well as the clutter from the environment. 
can be applied. Among several methods that can be designed to find the beamformer weights, we describe one next, where orthogonal weights can be constructed to cover the desired FoV, for a given array geometry. 
%After the detection of targets, the angle estimation can be further refined by applying other well-known methods. 
Adaptation on digital beamformer weights can be designed to give preference to an existing target direction, and to cover the other uncovered angles by the existing targets. Depending on $L$, the receiver beams can overlap for redundancy, and resilience to target movement. 
}

\textcolor{black}{
The procedure is as follows. Following the beam-space weight matrix computation method given in \cite{Anderson_ESP_1993}, we propose to select the orthogonal weights for the beamformer as the dominant eigenvectors of the covariance matrix given by 
\begin{equation}
    \mathbf{Q} = \int_{\theta_1}^{\theta_2} w(\theta) \mathbf{a}(\theta) \mathbf{a}^H(\theta) d\theta , 
    \label{eqn:arrayCovMat}
\end{equation}
where $\theta_1$ and $\theta_2$ denotes the boundary angles of the FoV and $w(\theta)$ denotes a weight function. In case of $M < L$ beams are desired, then $M$ dominant eigenvectors in $\mathbf{Q}$ can be used. One such design is illustrated later in Sec.~\ref{sec:simResults}. 
}

\subsection{Joint Radar and Communication} \label{subsec:jrc}
\textcolor{black}{
As mentioned earlier in Sec.~\ref{sec:datamodel}, we consider a JRC system in which the radar transmits data in the odd frames, and the targets communicate back in even frames. This simple protocol can avoid overlapping transmissions between two targets in close proximity. If the targets are distant, then they can be allowed to respond in all even frames. For simplicity, let the radar and the targets use the same codebook and coding rate. A suitable modulation and coding scheme can be used for data transfer. A practical Gaussian spherical code construction for Gaussian channel is described in Appendix~\ref{app:codeConstr}, which may be used to modulate the Gaussian pulses (or an optimally designed waveform) as described in the above section. When the targets transmit their modulated pulses, the following procedure is implemented to decode the bits and estimate the target parameters.
\begin{enumerate}[Step 1:]
    \item First, digital beamforming is done on the received data pulses, and the weights computed in Sec.~\ref{sec:beamWeights} is used for a suppression of inter-target interference in the angle domain. Note that this allows spectrum overlapping, which was one of the constraints in \cite{Cnaan-On_MTT_2015}, which limited the maximum data rate for the targets.
    \item The output of the beamformer is used to correlate with the baseband equivalent of the transmitter waveform.
    \item The correlation peaks identify the target distances. The correlator peak outputs corresponding to all such transmitted pulses are summed up to enhance the SNR and resilience in range detection. This gives inter-target interference suppression in the range domain.
    \item The correlator peak output for all targets are passed to the FFT in Doppler dimension to estimate the velocity, and the corresponding frequency is corrected on the correlator peak outputs. This gives inter-target interference suppression in the doppler domain.
    \item The resultant Doppler peak locations -- for every range correlator peak -- across all antennas are sent to the angle estimation block, which further refines the digital beam angles from which the range detection was obtained.
    \item The refined angles are used to apply beam weights on the range domain data, followed by the Doppler frequency correction to obtain the baseband signal corresponding to the data transmitted by the individual targets. 
    \item Conventional data demodulation and error correction are performed, including the matched filtering for the transmit pulse shape.
\end{enumerate}   }
\textcolor{black}{Thus, one can perform both the Radar target parameter estimation and data transmission using the same Transmit waveform, except that data is overriding on top of the transmit pulse envelope. Any incoherent sums used to improve SNR for range and Doppler frequency estimation remains unaffected by the code sent by the targets. 
}

\section{Bounds on Achievable Data Rates}   \label{sec:dataBounds}
\subsection{Mutual Information of the Forward Channel} \label{subsec:fc}
Consider the forward channel between the transmitter and $k^{\text{th}}$ target. Let $\mathbb{E}[a_i^2] = \sigma_x^2 = \pavg$. The reflected signal $w_{k,i}(t), i=1,\ldots,N$, from the target can be written as
\begin{align}
& w_{k,i}(t) \hspace{-0.05cm} = \hspace{-0.05cm} a_i \alphak \gamma_k \hspace{-0.1cm} \int_0^{t_k} \hspace{-0.15cm} g_k(t') x([t \hspace{-0.05cm}- \hspace{-0.05cm} i \ts - \hspace{-0.05cm} \tau_k \hspace{-0.05cm} - \hspace{-0.05cm} t']\mathbb{T}_k) dt' \nonumber \\
& ~~~~~~~~~~~~~~~~~~~~~~~~~~~~~~~~~~~~~~~~~ + j_{k,i}(t) + n_i(t),
\end{align}
where $j_{k,i}(t)$ is the interference for $k^{\text{th}}$ target due to a deliberate jammer. For the ease of analysis, we assume that $j_{k,i}(t)$ is a Gaussian random process with zero mean and variance $\sigjk$. However, the following analysis can be extended to other popular statistical models for clutter and jammer, e.g.~Gamma or Rayleigh, by suitable modifications. Now, the \textcolor{black}{signal-to-interference and noise ratio (SINR)} due to the reflected signal from the $k^{\text{th}}$ target \textcolor{black}{in the presence of other surrounding targets and jammers} can be written as 
\begin{equation}
\rhok = \frac{\alphak \gamma_k \sigma_x^2 \mathbb{G}_k}{\signk + \sigjk + \sigma_x^2 \sum_{j \ne k} \frac{\lambda^2 \mathbb{G}_j  \alpha_j \gamma_j }{16 \pi^2 (r_k-r_j)^2} },
\label{eqn:SNR_forwardCh}
\end{equation}
%\begin{equation}
%\rhokf = \frac{\left(\sigma_x \alphak \gamma_k\right)^2 %\mathbb{G}_k}{\sigma_j^2 + \signk} =  %\left[\frac{\sigma_x^2 \alphak^2}{\sigma_n^{(k)}^2}\right] %\left[\frac{\gamma_k^2 \mathbb{G}_k}{1 + %\left(\frac{\sigma_j^2}{\sigma_n^{(k)}^2} \right)} \right],
%\label{eqn:SNR_forwardCh}
%\end{equation}
where only the targets in close proximity to $k^{\text{th}}$ target are considered along with the corresponding free space pathloss, with $\lambda$ denoting the wavelength of the carrier. Moreover, 
\begin{align}
    \mathbb{G}_k \triangleq \int_{-B}^B \left|X_i\left(\frac{\omega}{\mathbb{T}_k}\right) G_k(\omega)\right|^2  d\omega
\end{align}
denotes a factor that determines the variance of the reflected signal due to target channel response  \cite{Papoulis_Book_1991}, and 
$\signk$ denotes the noise variance at the $k^{\text{th}}$ target's receiver. Note that \textcolor{black}{$\rho_k$ is calculated based on the reflected signal power due to the fact that it is proportional to the RCS of the target, and RCS captures the} maximum antenna aperture that is available at the target. 
%Hence, Assuming that the $N$-length code $\mathbf{c}$ is applied at the transmitter, recall that the differential entropy in the transmit waveform is given by $\frac{N}{2} \log(2 \pi e \sigma_x^2)$ nats. Additional scaling factors due to path loss and reflection coefficients only affect the variance of the Gaussian process, and can be incorporated into the SNR defined in \eqref{eqn:SNR_forwardCh}. 
\textcolor{black}{As an illustrative example, consider a target with no interference from jammers, i.e.~$\sigma_j^2=0$, $\mathbb{G}_k=1$, $\gamma_k = \kappa_k \mathbf{A}_k$, $\alphak = \frac{1}{4 \pi r_k^2}$ and ${\rhot}_k = \frac{\sigma_x^2}{ \signk}$, where $\mathbf{A}_k$ denotes the surface area of the target, $0 \le \kappa_k \le 1$ denotes the aperture efficiency. Then, the SINR at the $k^{\text{th}}$ target on the forward channel simplifies to}
\begin{equation}
\rhokf = 
\frac{ {\rhot}_k  \kappa_k \mathbf{A}_k}{4 \pi r_k^2 \left(1 +  \frac{\sigjk}{\signk} + {\rhot}_k  \sum_{j \ne k} \frac{\lambda^2 \mathbb{G}_j  \alpha_j \gamma_j }{16 \pi^2 (r_k-r_j)^2} \right)},
\label{eqn:SNR_forwardCh2}
\end{equation}
\textcolor{black}{which is equivalent to the SNR of the signal at the $k^{\text{th}}$ target. Hence, a bound on} the mutual information in the forward channel between the transmitter and $k^{\text{th}}$ target, $I(X,W_k)$, can be computed in a straightforward manner using the linear model and AWGN channel model as shown in Fig.~\ref{fig:target_channel_model} \cite{TomCover_Book_1991}. That is,
%\begin{align}
%& I(X,W_k) \ge \frac{N}{2} \log \left(1 \hspace{-0.1cm} + \hspace{-0.1cm} \frac{(\sigma_x \alphak \gamma_k)^2 \mathbb{G}_k}{\sigma_j^2 \hspace{-0.1cm} + \hspace{-0.1cm} \signk \hspace{-0.1cm} + \hspace{-0.1cm} \sigma_x^2 \sum_{j \ne k} \sbtot  \mathbb{G}_j \left[\frac{\alpha_j \gamma_j}{\alpha_f} \right]^2} \right) \nonumber \\
%& \phantom{I(X,W_k)} =   \log \left(1+\rhokf \right)^{\frac{N}{2}},
%\label{eqn:forwardRate}
%\end{align}
\begin{align}
& I(X,W_k) \ge \frac{N}{2} \log \left(1 \hspace{-0.1cm} + \hspace{-0.1cm}   \frac{\alphak \rhot \gamma_k \mathbb{G}_k}{1 + \frac{\sigjk}{\signk} + {\rhot}_k \sum_{j \ne k} \frac{\lambda^2 \mathbb{G}_j  \alpha_j \gamma_j }{16 \pi^2 (r_k-r_j)^2} } \right) \nonumber \\
& \phantom{I(X,W_k)} =   \log \left(1+\rhokf \right)^{\frac{N}{2}}.
\label{eqn:forwardRate}
\end{align}
Therefore, \eqref{eqn:forwardRate} gives a lower bound on the data rate that can be communicated from the radar transmitter to $k^{\text{th}}$ target.

\subsection{Mutual Information of the Reverse Channel} \label{subsec:rc}
Similar to the derivation in Sec.~\ref{subsec:fc}, we can derive a lower bound on the mutual information for the reverse channel between each passive target $k$ and the receiver as
\begin{align}
& I(X;Z_k) \hspace{-0.05cm} \ge \hspace{-0.05cm} \frac{N}{2} \log \hspace{-0.1cm} \left( \hspace{-0.1cm} 1  \hspace{-0.1cm} + \hspace{-0.1cm} \frac{\alphak^2 \rhot \gamma_k \mathbb{G}_k}{1+\rhot \hspace{-0.1cm} \left\{\underset{j \ne k}{\sum} \sbtot  \alpha_j^2 \gamma_j \mathbb{G}_j \right\} } \hspace{-0.1cm} \right) \nonumber \\ 
& \phantom{I(X;Z_k)} \hspace{-0.05cm} = \hspace{-0.05cm} \log \left(1 + \rhokr \right)^{\frac{N}{2}},
\label{eqn:reverseRate}
\end{align}
where $\rhokr$ is the SINR at the receiver from the reverse channel, \textcolor{black}{and $\sbtot$ denotes the interference suppression obtained by the three dimensional filtering, given in \eqref{eqn:sb_total}.}\footnote{\textcolor{black}{We ignore the interference due to jammers to derive \eqref{eqn:reverseRate}, under the assumptions that (a) There are no jammers near the radar receiver, (b) the jamming signals are significantly attenuated at the radar receiver, (c) the jamming signals get filtered out by the beamforming procedure.}} The expression in \eqref{eqn:reverseRate} yields a lower bound on the achievable data rate between the $k^{\text{th}}$ passive target and the radar source. Recall that the target is assumed to be a passive reflector, and only changes its reflection coefficients to perform modulation on the reflected signal from the targets. For this reason, the strength of the reflected signal is taken as the received signal power in the reverse channel. \textcolor{black}{Note the similarity between the expressions in \eqref{eqn:forwardRate}
and \eqref{eqn:reverseRate}, except for the additional $\alphak^2$ in the numerator for the reverse-channel to account for the additional pathloss and $\sbtot$ term in the denominator which provides the 3-D filtering owing to the separation of the targets in 3-D space namely, range, velocity and angle of arrival.}
%In the next section, we consider the radar performance and derive lower bounds on the limits on the MSE of the parameter estimates, namely range, velocity and azimuth.

\section{Results and Discussion} \label{sec:simResults}
\textcolor{black}{
In this section, we illustrate the utility of various design methods discussed in the previous sections, using examples through numerical techniques. 
}

\subsection{Optimization of Transmit Waveform}
\begin{figure*}[!t]
\centering
\subfigure[]{
    \includegraphics[width=0.31\textwidth]{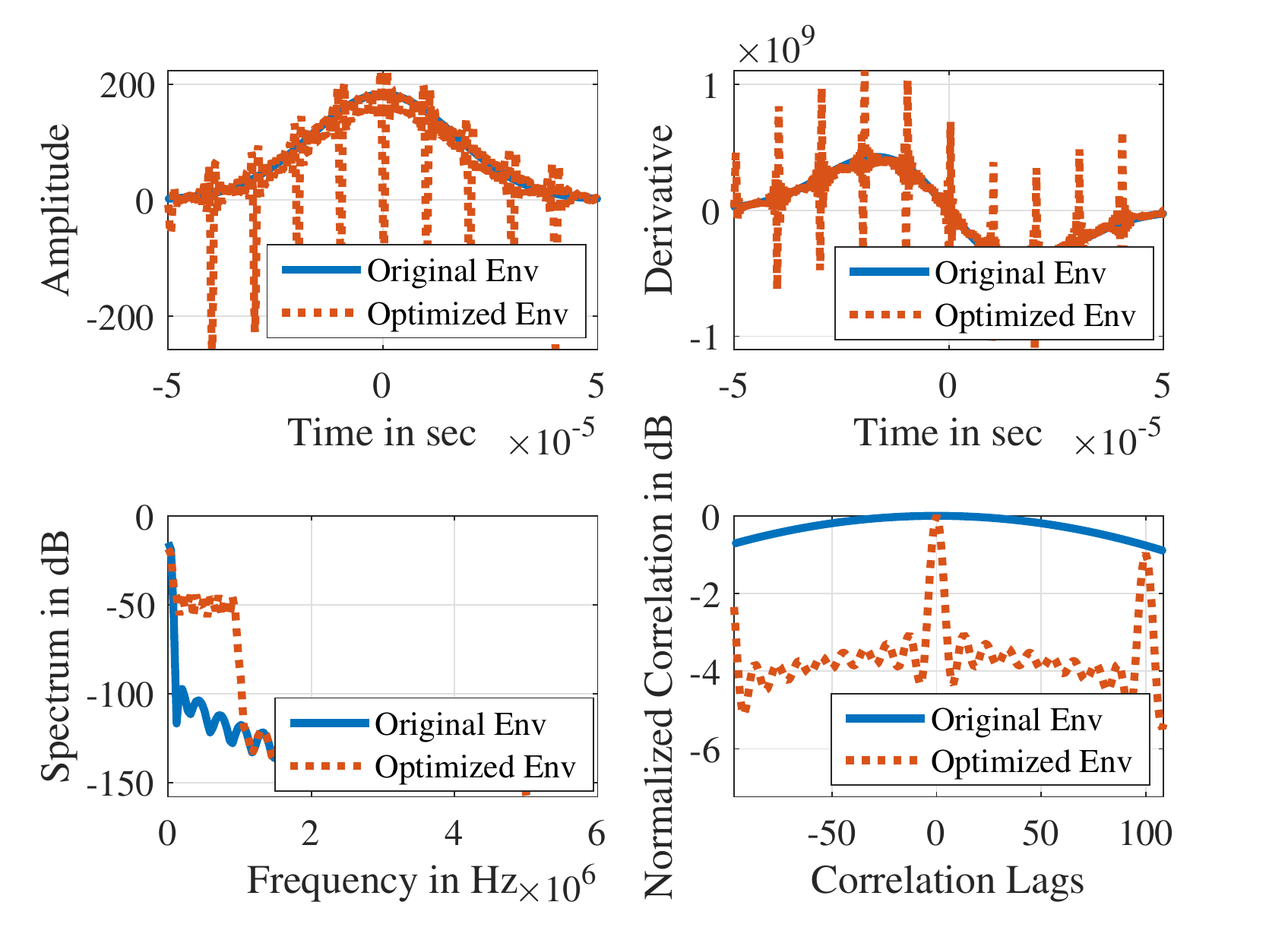}
    \label{fig:gaussProperties1MHz}
}
\subfigure[]{
    \includegraphics[width=0.31\textwidth]{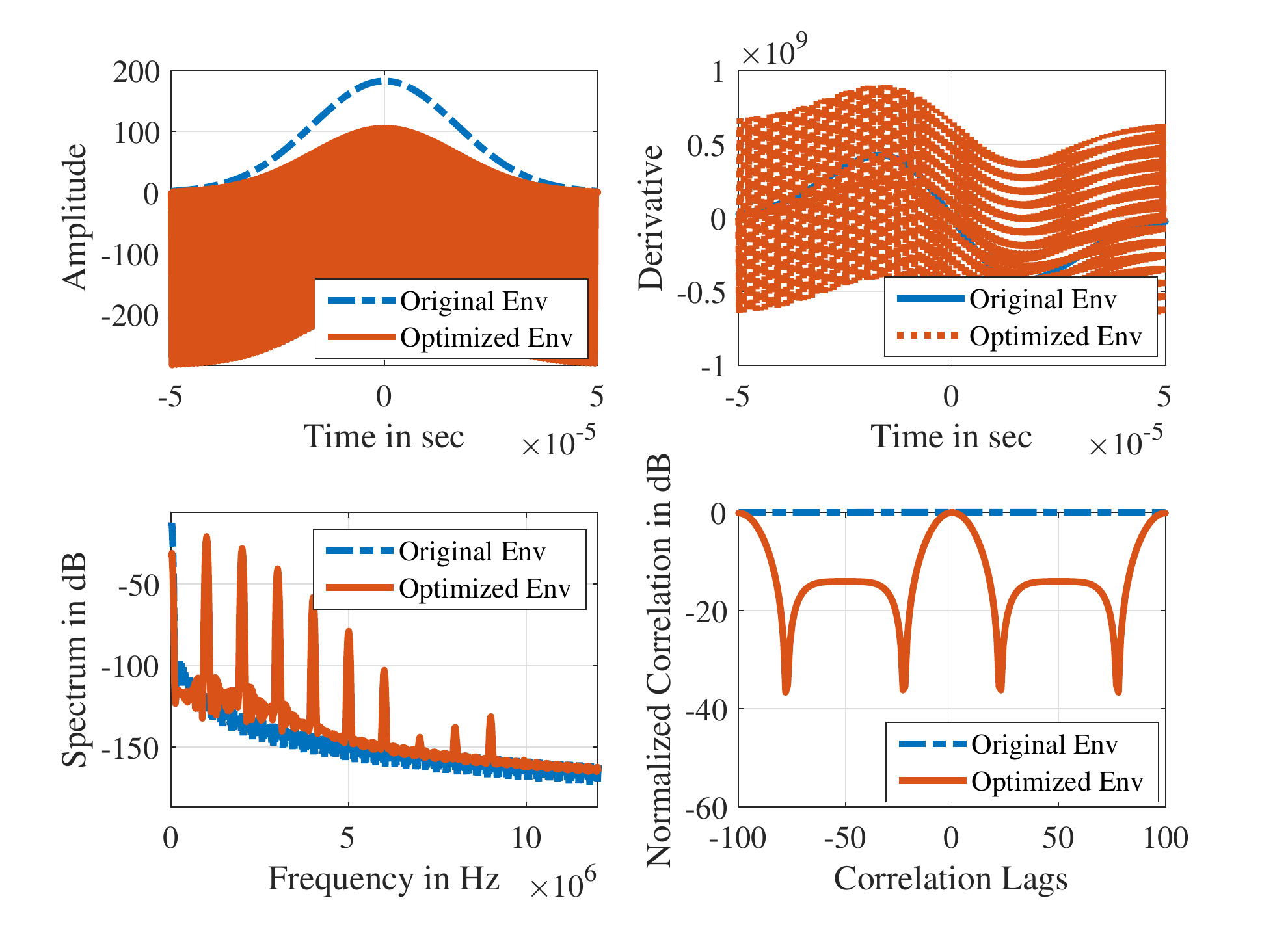}
    \label{fig:gaussXcorr1MHz}
}
\subfigure[]{
    \includegraphics[width=0.31\textwidth]{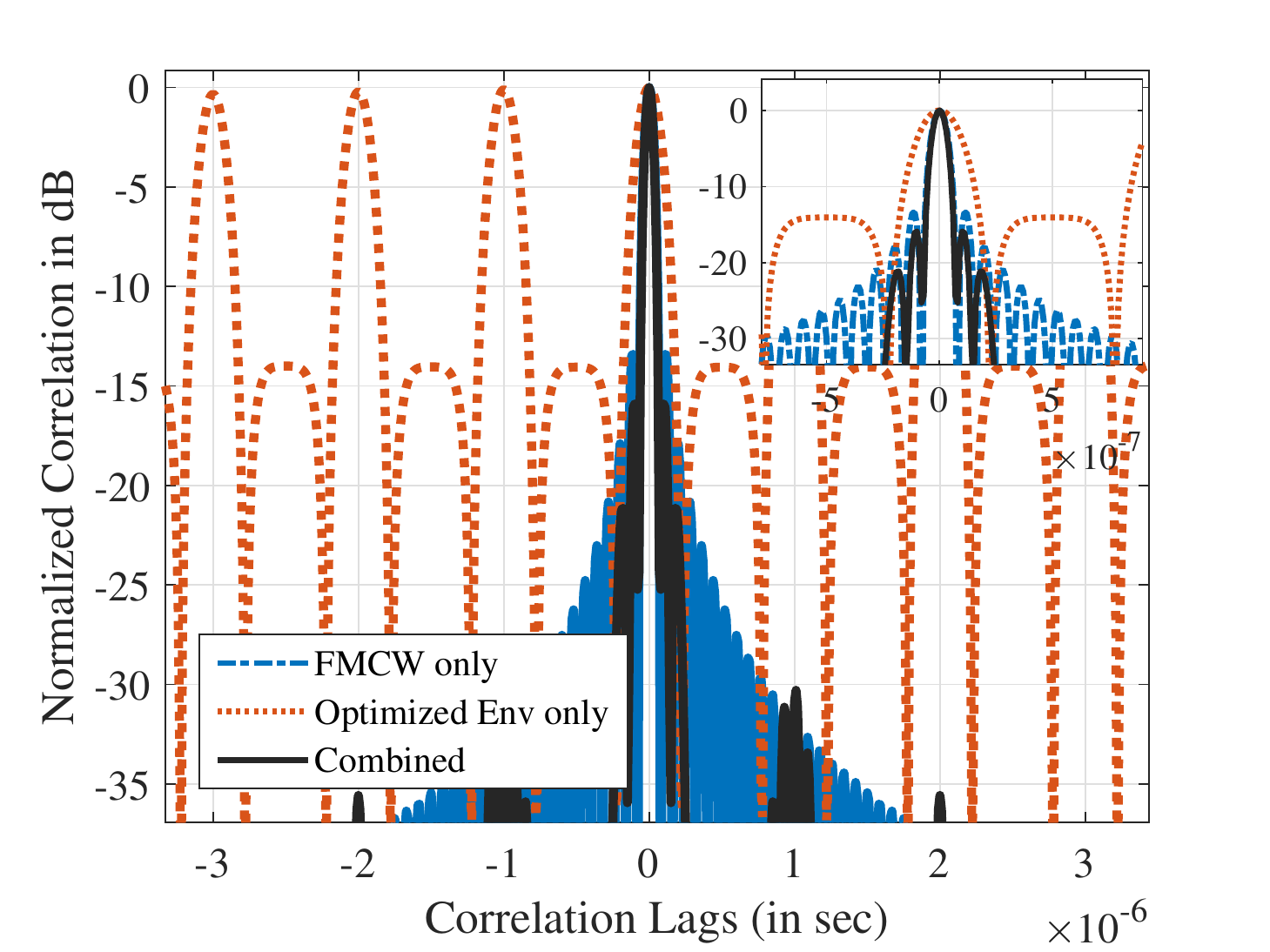}
    \label{fig:gaussXcorr10MHz}
}
\vspace{-0.3cm}
\caption{\textcolor{black}{(a) Properties of the optimized Gaussian pulse waveform with $1$ MHz bandwidth constraint, (b) Properties of the optimized Gaussian pulse waveform with $10$ MHz bandwidth constraint, and (c) Autocorrelation of the optimized envelope and FMCW modulated with the optimized envelope.}}
\label{fig:subfigureExample}
\hrulefill
\end{figure*}

\textcolor{black}{
In this subsection, we demonstrate the optimal waveform design obtained by the CoV-based optimization. To begin with, the waveform was chosen to be Gaussian with} $T_s = 100~\mu$s. \textcolor{black}{In general, the choice can be any waveform listed in Table~\ref{tab:waveforms}. The cost function chosen was the sum of correlation lags from $2$ to $20$ units. Reduction of the cost function was carried out over $200$ iterations. Figures~\ref{fig:gaussProperties1MHz}-\ref{fig:gaussXcorr10MHz} show the comparison of various properties of the designed optimal envelope 
%by minimizing the chosen cost function which also minimizes the normalized correlation lags. The normalized correlation lags denote the interference one targets exerts on its neighbor. 
Note that the optimal envelope converges to a series of impulse functions. Depending on the bandwidth expansion allowed in the new envelope function, the correlation lags were suppressed further. That is, the correlation suppression varies directly with the envelope bandwidth. This bandwidth expansion constraint was added as part of the iterations explained Section~\ref{subsec:txwaveform}. Figure~\ref{fig:gaussXcorr10MHz} shows correlation lags between $100$ ns and $1$ $\mu$s, which are lower than that observed over the unmodulated FMCW carrier. In this example, the FMCW carrier has a bandwidth of $10$ MHz, which can resolve targets separated by $100$ ns. Thus, the $100$ ns limitation in the optimal waveform comes from the carrier rather than from the envelope. This demonstrates that the optimizing the TAF functions reduces the interference between two targets, as discussed in Sec.~\ref{sec:EstBounds}}.

% \begin{figure*}[!t]
% \centering
% \begin{minipage}{0.33\textwidth}
%   \centering
% \includegraphics[width=0.33\textwidth]{gaussian200Iter_3to20_lags_xcorr_compare_optimized_1MHz.pdf}
% %\subcaption[first caption.]{First}\label{fig:1a}
% \caption{First}\label{fig:1a}
% \end{minipage}%
% \begin{minipage}{0.33\textwidth}
%   \centering
% \includegraphics[width=0.33\textwidth]{gaussian200Iter_3to20_lags_xcorr_compare_optimized_10MHz.pdf}
% %\subcaption[second caption.]{Second}\label{fig:1b}
% \caption{Second}\label{fig:1b}
% \end{minipage}%
% \begin{minipage}{0.33\textwidth}
%   \centering
% \includegraphics[width=0.33\textwidth]{gaussian200Iter_3to20_lags_xcorr_compare.pdf}
% %\subcaption[third caption.]{Third}\label{fig:1c}
% \caption{Third}\label{fig:1c}
% \end{minipage}
% %\caption{General caption.} \label{fig:1}
% \hrulefill
% \end{figure*}

\subsection{Receive Beamforming}
\begin{figure}[!t]  
  \begin{center}
  %\vspace{-1in}
  \includegraphics[width=0.45\textwidth]{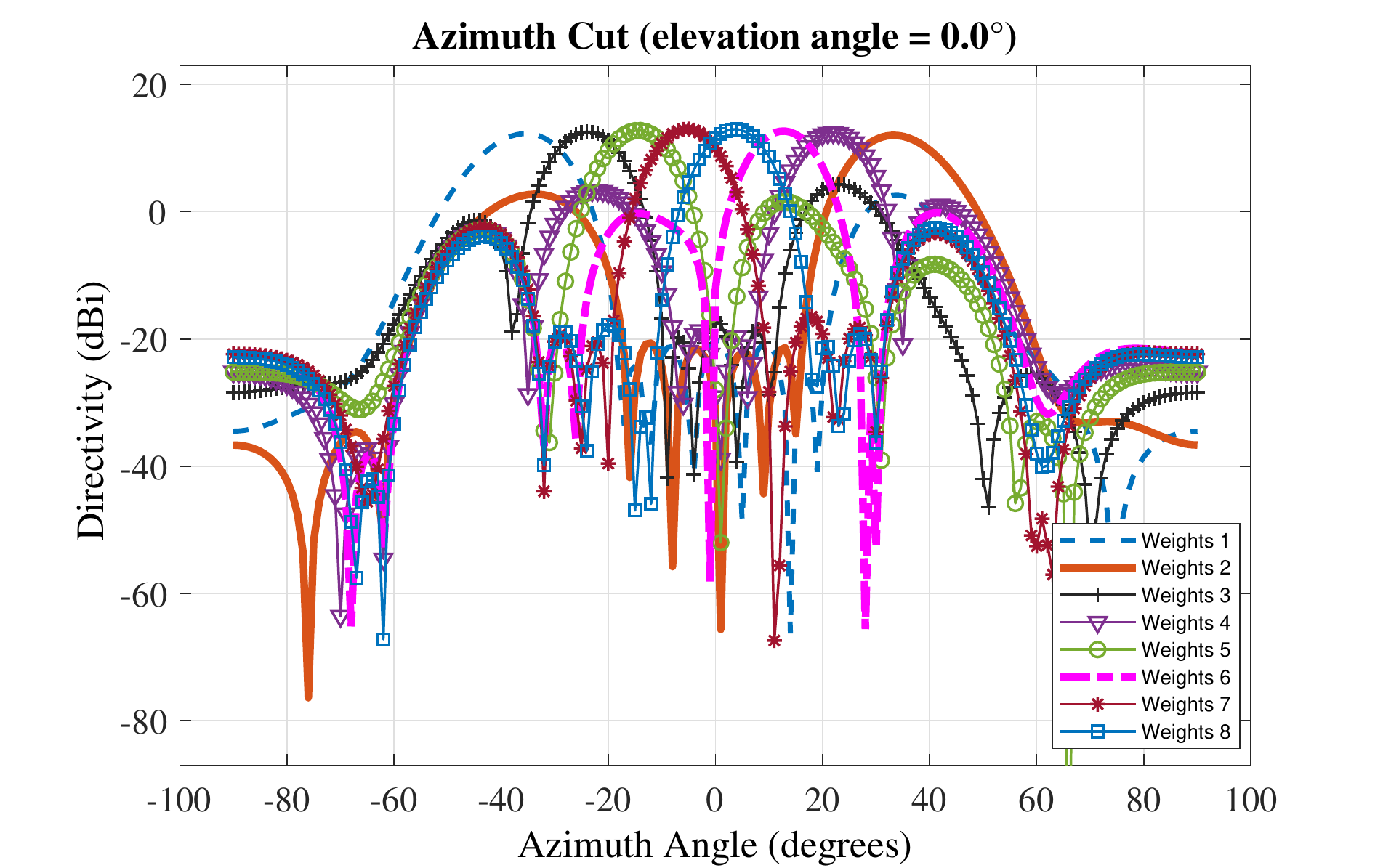}
 \end{center}
%\vspace{-1in}
\caption{\textcolor{black}{Beamformer output for various weights designed with Taylor window weighting.}}
\label{fig:taylorWinBeams}
\end{figure}

\textcolor{black}{
Consider an $L=12$ element uniform linear array (ULA) of antennas. We intend to design the digital beamformer weights whose field of view (FoV) is $\pm 45$ degrees. Figures shows the performance of the designed beamformer.
%, in comparison with the performance of a design with standard array steering vectors. 
Note that the standard steering vectors are not orthogonal and there is no direct way to incorporate the weight function for selecting a particular sector or FoV. Figure~\ref{fig:taylorWinBeams} shows the array response for a set of $8$ weights after applying Taylor window array tapering. By design, the array weights are orthogonal -- before applying the array taper, and one can design upto $L$ such vectors from the eigenvectors of the covariance matrix for the given array geometry. Note that the side-lobes and main beam width can be altered using different array tapering values. These weights were obtained for $\theta_1 = -50$ degrees and $\theta_2 = 48$ degrees. This asymmetry is deliberately created to produce complex eigenvectors. As observed, the sidelobes are suppressed by more than $30$ dB in the Taylor window tapered weights case.
}

%\begin{figure}[!t]  
%  \begin{center}
%  \vspace{-1in}
%  \includegraphics[width=3.35in]{a%rrayResp_KaiserWin_m50p48deg.pdf}
% \end{center}
%\vspace{-1in}
%\caption{Beamformer output for various %weights designed with Kaiser window %weighting.}
%\label{fig:kaiserWinBeams}
%\end{figure}

\subsection{Parameter Estimation in JRC}
\textcolor{black}{
A JRC system was implemented using the Radar toolbox in Matlab, with the parameters given in Table~\ref{tab:jrcSimParams}. Two targets, one moving slow and one moving fast, both moving away from the radar, were chosen. 
%An FMCW radar with $100$ MHz chirp bandwidth with Gaussian pulses of duration $100$~ $\mu$s was implemented, where $512$ chirps were transmitted per frame. $200$ frames were simulated with and without transmitting the data. 
The $(32,8)$ Gaussian codebook designed in Appendix~\ref{app:codeConstr} was used for transmission. It can be seen that in both cases, the performance of the radar is almost identical. The error in the measured position is mainly due to the finite angle resolution due to the $12$ element ULA used and simple FFT processing was used to detect all three parameters, range, velocity and angle of arrival. The received signal after pulse compression was sampled at $40$~MHz sampling rate with $4096$ point range FFT and $512$ point Doppler and angle FFTs taken for computing the radar parameters. The target detection was done in $3$ stages, first two using order statistics-based range and Doppler detection, and the final based on a 2D centroid detection in the range-Doppler domain. Thresholds were chosen to achieve a constant false-alarm rate. 
%The details of the radar implementation is beyond the scope of this paper. 
}

\begin{figure}[!t]  
  \begin{center}
%  \vspace{-1.0in}
  \includegraphics[width=7cm, height=6cm, trim={3.5cm 0 4cm 0},clip]{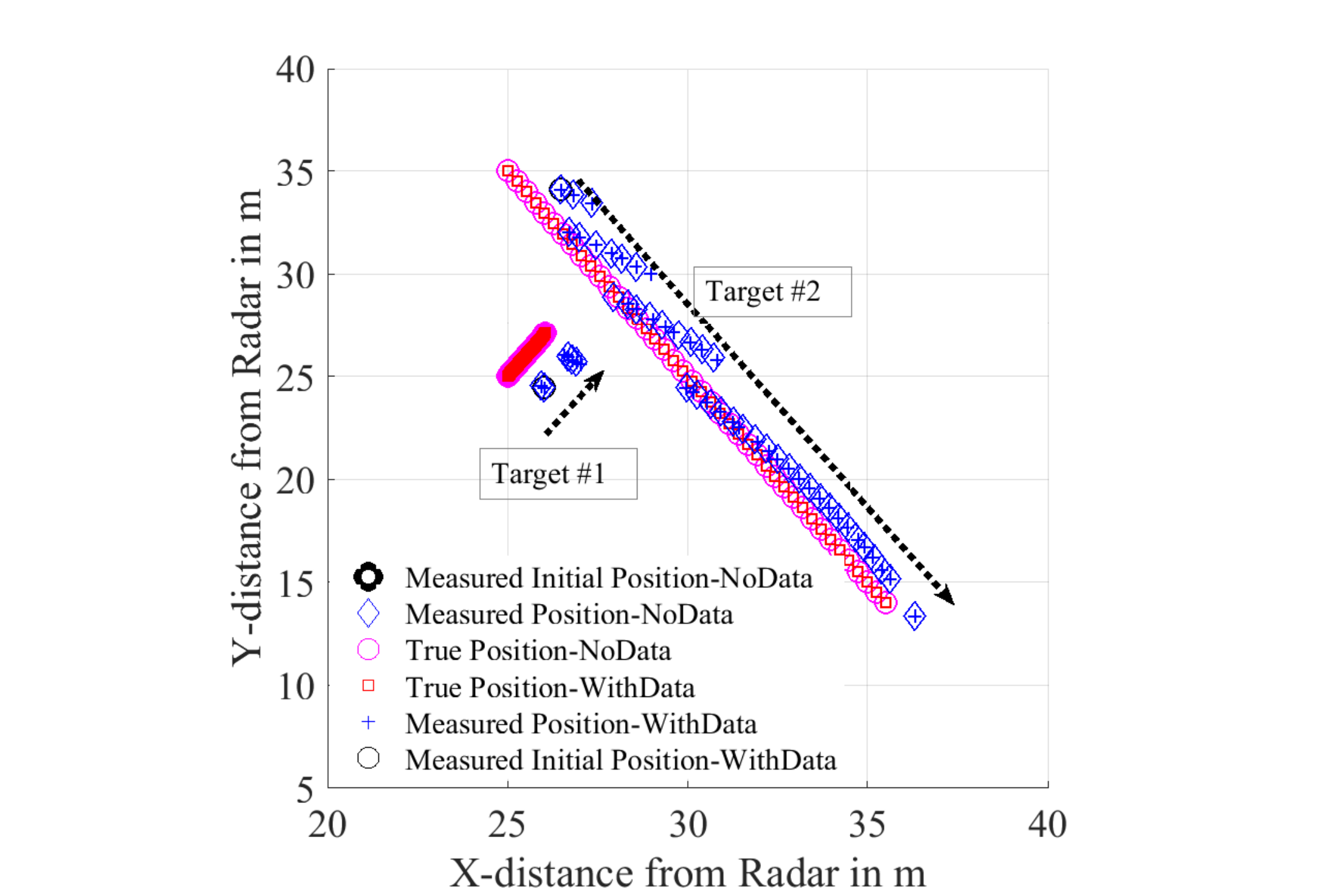}
 \end{center}
%\vspace{-1.5in}
\caption{\textcolor{black}{$24$ GHz FMCW radar performance with and without data transmission using Gaussian envelope.}}
\label{fig:compareRadarPerf}
\end{figure}

\textcolor{black}{
In the case where data transmission was done, the data was received with zero bit error rate for both targets since SINR is high ($> 4$ dB even at the farthest radial distance of $43$ m). Just to compare the effect of data transmission on the Radar performance, no clutter was added and target RCS was kept constant i.e., no random fluctuations such as Swerling model was used. But practical radar receiver impairments such as receiver noise, automatic gain control, and ADC quantization were added in this study. Note that no tracking and smoothing algorithms such as Kalman filtering were applied on the data. The raw measured target position parameters are plotted in  Figure~\ref{fig:compareRadarPerf} after adjusting for the antenna height. This demonstrates that if one can design radar transmit waveforms such that the sidelobe leakage of one target onto another (Refer to Eqn. \eqref{eqn:sb_total}) is minimized, then we can achieve joint Radar communication with no loss in performance of parameter estimation as well as data transmission system simultaneously. Moreover, one can achieve the performance bounds given in Eqns. \eqref{eqn:reverseRate} and  \eqref{eqn:pd_range0}-\eqref{eqn:pd_angle0}. Rigorous proof of achievability is not discussed here, and will be taken up as a future work.
}

\begin{center}
\begin{table}
    \centering
    \begin{tabular}{|c||l|c|c|}
    \hline 
        Sl. & Parameter & Value   & Unit \\
        \hline \hline
        1   & No. of transmit antennas & $1$ & - \\    \hline 
        2 & No. of receive antennas & $12$ & - \\  
        \hline
        3 & Array geometry & ULA & - \\  
        \hline
        4 & Antenna spacing & $0.0078$ & m \\  
        \hline
        5 & Antenna height & $8$ & m \\  
        \hline
        6 & Antenna grazing angle & $6.38$ & degrees \\  
        \hline
        7 & Antenna beamwidths & $(-45,45)$ & Azimuth deg \\
        &                      & $(-6,6)$ & Elevation deg \\
        \hline
        8 & FMCW Carrier bandwidth & $100$ & MHz \\
        \hline
        9 & Num.~of chirps per frame & $512$ & - \\
        \hline
        10 & Chirp time, $T_s$ & $100$ & $\mu$s \\
        \hline
        11 & Envelope bandwidth & $100$ & KHz \\
        \hline
        12 & Carrier frequency & $24$ & GHz \\
        \hline
        13 & Forward error correction code & $(32,8)$  & - \\
        \hline
        14 & Target data rate & $2500$ & bits/s \\
        \hline
        15 & Transmit pulse envelope & Gaussian & - \\
        \hline
        16 & RCS of targets & $1$ & m$^2$ \\
        \hline
        17 & Target velocities & $(0.5,1)$ & m/s \\
        &   ($v_x,v_y$)        & $(5,-10)$ & m/s \\
        \hline
        18 & Target start positions & $(25,25)$ & m \\
        &   ($r_x,r_y$)        & $(25,35)$ & m \\
        \hline
    \end{tabular}
    \caption{\textcolor{black}{Parameters for the simulation of a JRC system.}}
    \label{tab:jrcSimParams}
\end{table}
\end{center}
%\end{figure*}

\subsection{Data Rate Bounds in JRC}
\textcolor{black}{
To demonstrate the utility of the data rate expressions for the forward and reverse channel (Eqns. \eqref{eqn:forwardRate} and \eqref{eqn:reverseRate}), numerical simulation was performed to compute the data rate bounds in the forward and reverse channel for the above 24 GHz radar (but with 1 antenna at both ends and no error correction code applied) where $\rhot$ and ${\rhot}_k$ are assumed to be $100$ dB. We assumed two static, point reflection targets with 1 sq.meter RCS area and no jammer. The two targets and the radar make an right angle triangle with different acute angle subtended at the radar depending on the distance between the two targets. Moreover it is assumed that, $\gamma_k = \mathbb{G}_k = 1$  and $\alpha_k = \frac{1}{4 \pi r_k^2}$ for all targets. The data rate is plotted as a function of distance from the radar. 
%Single antenna is used for both transmit and receive. 
Note that, the distance of target 2 is always larger than that of target 1 by a fixed amount for one simulation.
}

\begin{figure}[!t]  
  \begin{center}
  %\vspace{-1.5in}
  \includegraphics[width=0.5\textwidth]{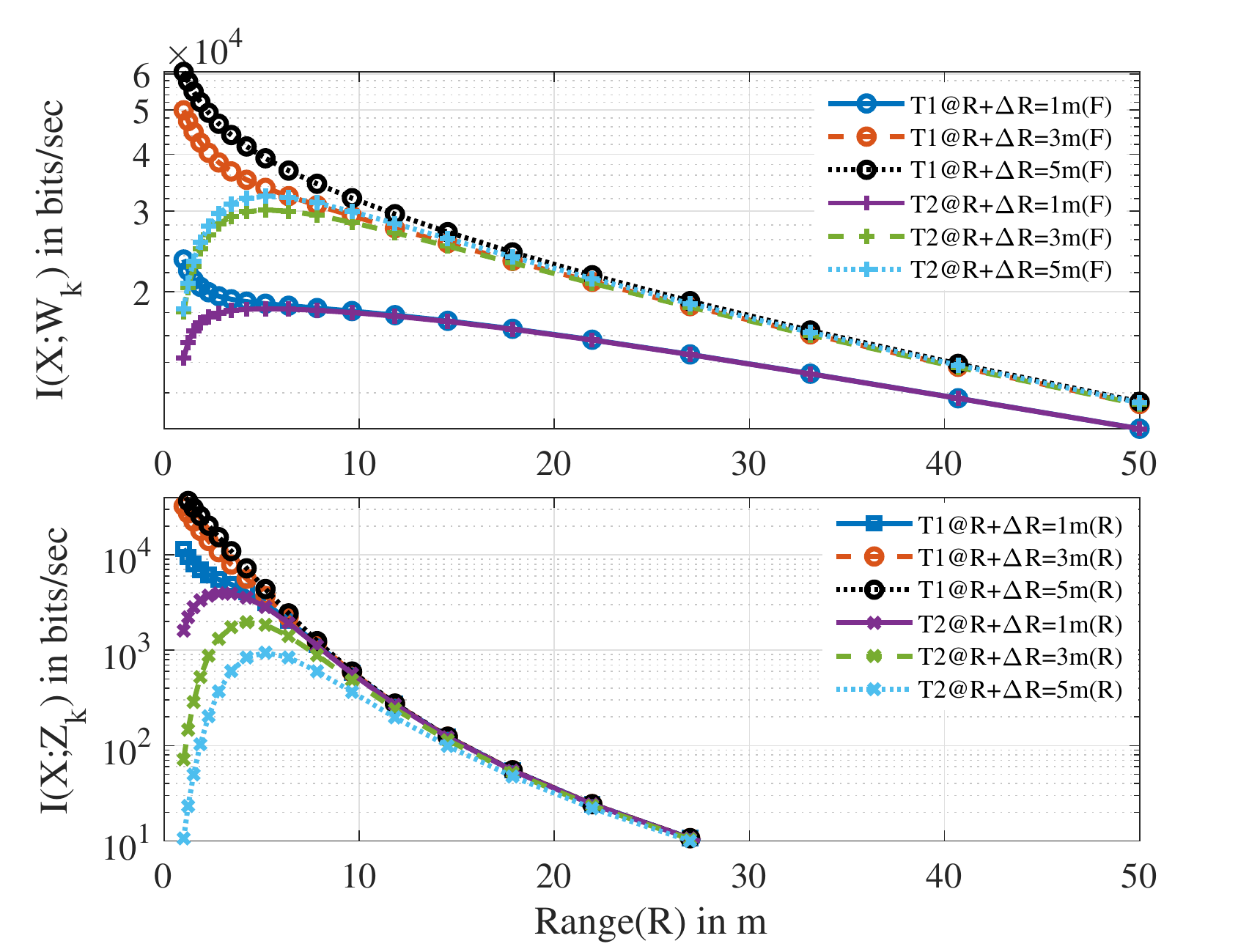}
 \end{center}
%\vspace{-1.5in}
\caption{\textcolor{black}{Data rate bounds for the forward and reverse channels in a $24$ GHz JRC radar.}}
\label{fig:fwdRevDataRate}
\end{figure}

\textcolor{black}{
Figure~\ref{fig:fwdRevDataRate} shows the data rate bounds for both channels for 3 different distances between the targets. It can be observed that the forward channel rate $I(X,W_k)$ for the target 1 reduces with an increase in range. This reduction is sharp at small range compared to the linear reduction at high range values. This is largely due to the interference from target 2 causing additional loss in the received SINR at low range. Note that the data rate for target 1 is higher when target 2 is farther away. On the other hand, data rate for target 2 first increases and then starts to decrease, due to the reduction in the interference from target 1 at close ranges. Further, it is observed that the data rates increase as the separation between the targets increase. Both targets converge to nearly equal data rates as their distance from the radar increases. However, a minor difference in rates exists due to the difference in their distances from radar as well as distance between them. Looking at the reverse channel rate $I(X,Z_k)$, we observe that the rate for target 1 decreases with distance and is lower than $I(X,W_k)$. Data rate for target 2 improves with range, and decreases with an increase in separation between them. Due to non-availability of multiple antennas at the receiver, no spatial filtering can be designed except for $SB_{\text{range}}$. Moreover, note that the drastic reduction in $I(X,Z_k)$ with distance, due to  higher (round trip) pathloss on the reverse channel compared to the forward channel case.
}

\section{Conclusion and Future Work} \label{sec:Conc}
\textcolor{black}{In this paper, a novel analytical approach using a linear channel target response model was employed to derive lower bounds for data rates and parameter estimation errors. This framework is valid in the context of a wideband JRC with semi-passive targets, for any given array geometry. The radar transceiver was assumed to not only detect the targets, but also to exchange information with them. The targets use controlled backscattering of the transmitted signal for data transfer, and receive the data using active components. It was shown that the proposed inter-target interference suppression improves the performance of the JRC system on the reverse channel. Moreover, it was observed that the forward channel data communication does not impact the parameter estimation bounds. Furthermore, the data bounds were shown to be achievable if and only if the TAFs are optimized by proper selection of the transmit waveforms. Using CoV, an optimal transmit waveform was designed to minimize the range TAF. By optimizing three different TAFs, it was shown that one can achieve a desired performance for each of the parameters. The proof of achievability of these bounds are a part of the future work. Additionally, design of practical codes to achieve these bounds is an interesting research area.}

\begin{appendix}
\subsection{Target Ambiguity Functions} \label{app:taf}
The TAFs are generalized versions of the conventional ambiguity functions, which arise naturally in the computation of CRBs. This requires computation of second order partial derivatives of logarithm of the joint PDF function. For the radar parameter estimation, the partial derivative with respect to the parameters can be computed using
\begin{align}
\frac{\partial \log f_z(\mathbf{z}_i)}{\partial r_m} = \left[ \frac{\partial \log f_z(\mathbf{z}_i)}{\partial \left(\mathbf{z}_i-a_i \mathbf{X}_i \Gamma_i \right)} \right]^T 
\frac{\partial \left(\mathbf{z}_i-a_i \mathbf{X}_i \Gamma_i \right) }{\partial r_m}.
\end{align}
Upon simplification, we get
\begin{align}
& \frac{\partial \log f_z(\mathbf{z}_i)}{\partial r_m} \hspace{-0.1cm}  = \hspace{-0.1cm} -\frac{1}{2} \left[ 2\left(\mathbf{z}_i \hspace{-0.1cm} - \hspace{-0.1cm} a_i \mathbf{X}_i \Gamma_i \right)^T \mathbf{R}_{n}^{-1} \right] \frac{\partial \left(\mathbf{z}_i-a_i \mathbf{X}_i \Gamma_i \right) }{\partial r_m}.
\end{align}
Further, it can be shown that
\begin{align}
& \left(\frac{\partial\ \mathbf{X}_i}{\partial\ r_m} \right)  =  \frac{4 \pi^2 \mathbf{T}_m^2 \gamma_m \sum_j \zeta_{mj} \delta(t-t_{mj})}{\beta^2\ c} \left[  \mathbf{x}_m  \circ \mathbf{t}_m \right],
\end{align}
\begin{align}
& \left(\frac{\partial\ \mathbf{\Gamma}_i}{\partial\ r_m} \right)  = \left[0, \ldots, 0, \gamma_m \sum_j \zeta_{mj} \delta(t-t_{mj}) \right. \nonumber \\ 
& ~~~~~~~~~~~~~~~~~~~~~~~~~ \left. \{ -2e\ \epsilon^2\ r_m^{-2e-1} \}, 0, \ldots, 0 \right]^T,
\end{align}
\begin{align}
& \frac{\partial\ \log f_z(\mathbf{z}_i)}{\partial\ r_m} \stackrel{(a)}{\approx}  a_i \frac{4 \pi^2 \mathbf{T}_m^2\ \gamma_m \sum_j \zeta_{mj} \delta(t-t_{mj})\ \epsilon^2 r_m^{-2e}}{\beta^2\ c} \nonumber \\
& ~~~~~~~~~~~~~~~~~~~~~~~~~~ \left(\mathbf{z}_i-a_i \mathbf{X}_i \Gamma_i \right)^T \mathbf{R}_{n}^{-1} \left[  \mathbf{x}_m  \circ \mathbf{t}_m \right] \nonumber
\end{align}
\begin{align}
& \frac{\partial^2\ \log f_z(\mathbf{z}_i)}{\partial\ r_m \partial\ r_n}   =  a_i \frac{4 \pi^2 \mathbf{T}_m^2\ \gamma_m \sum_j \alpha_{mj} \delta(t-t_{mj})}{\beta^2\ c} \nonumber \\
& ~~~~~~~~~~~~~~ \frac{\partial}{\partial\ r_n}\left\{ \left(\mathbf{z}_i-a_i \mathbf{X}_i \Gamma_i \right)^T \mathbf{R}_{n}^{-1} \left[  \mathbf{x}_m  \circ \mathbf{t}_m \right] \right\},
\end{align}
\begin{align}
& \frac{\partial}{\partial\ r_n}\left\{ \left(\mathbf{z}_i-a_i \mathbf{X}_i \Gamma_i \right)^T \mathbf{R}_{n}^{-1} \left[  \mathbf{x}_m  \circ \mathbf{t}_m \right] \right\} \nonumber \\
& ~~~~~~~~~~~ = \left(\mathbf{z}_i-a_i \mathbf{X}_i \Gamma_i \right)^T \mathbf{R}_{n}^{-1} \underbrace{\frac{\partial}{\partial r_n} \left[  \mathbf{x}_m  \circ \mathbf{t}_m \right]}_{=0} + \nonumber \\
& ~~~~~~~~~~~~~~~~~~~ \frac{\partial}{\partial r_n} \left(\mathbf{z}_i-a_i \mathbf{X}_i \Gamma_i \right)^T \mathbf{R}_{n}^{-1} \left[  \mathbf{x}_m  \circ \mathbf{t}_m \right],
\end{align}
where the approximation denoted by $(a)$ is obtained by ignoring the derivative of $\mathbf{\Gamma}_i$ with respect to $r_m$, since the first term dominates the second. 
\begin{figure*}
\begin{align}
\frac{\partial^2\ \log f_z(\mathbf{z}_i)}{\partial\ r_m \partial\ r_n}     
 =  -a_i^2 \frac{16 \pi^4 \mathbf{T}_m^4\ \gamma_m \gamma_n \mathbf{\zeta}_{m}  \mathbf{\zeta}_{n} \epsilon^4 r_m^{-2e} r_n^{-2e} }{\beta^4\ c^2} \left[  \mathbf{x}_n  \circ \mathbf{t}_n \right]^T \mathbf{R}_{n}^{-1} \left[  \mathbf{x}_m  \circ \mathbf{t}_m \right].
\label{eqn:eqn18}
\end{align}
\hrulefill
\end{figure*}
The final expression for $\frac{\partial^2\ \log f_z(\mathbf{z}_i)}{\partial\ r_m \partial\ r_n} $ is given in \eqref{eqn:eqn18} at the top of next page. It can be observed that $-\mathbb{E}\left[ \frac{\partial^2\ \log f_z(\mathbf{z}_i)}{\partial\ r_m \partial\ r_n} \right] = 0$, since $\mathbb{E}[\gamma_m \gamma_n] = 0$.

\subsubsection{TAF in Range}
Now, consider the function, 
\begin{equation} 
\mathcal{A}_r(r_m,r_n) \triangleq \mathbb{E} \left( \left[  \mathbf{x}_n  \circ \mathbf{t}_n \right]^T \mathbf{R}_{n}^{-1} \left[  \mathbf{x}_m  \circ \mathbf{t}_m \right] \right) , 
\label{eqn:TAF_ran}
\end{equation}
which measures the average weighted inner product between the two derived vectors, 
$\left[  \mathbf{x}_n  \circ \mathbf{t}_n \right]$ and $\left[  \mathbf{x}_m  \circ \mathbf{t}_m \right]$ where the term $\mathbf{x}_n  \circ \mathbf{t}_n$ is proportional to the derivative of $\mathbf{x}_n$ . Here, the expectation is taken over all possible differences in the range of two targets except for $r_m = r_n$. That is, the expectation can be computed as shown in \eqref{eqn:TAF_ran1}, 
\begin{figure*}
\begin{align} 
\mathcal{A}_r(r_m,r_n) = \int_{r_m=r_{min}}^{r_{max}} \ \int_{r_n=r_{min},r_n \ne r_m }^{r_{max}}   \left[  \mathbf{x}_n  \circ \mathbf{t}_n \right]^T \mathbf{R}_{n}^{-1} \left[  \mathbf{x}_m  \circ \mathbf{t}_m \right] f_{R_m, R_n}(r_m,r_n) dr_n dr_m  , 
\label{eqn:TAF_ran1}
\end{align}
\hrulefill
\end{figure*}
where $f(r_m, r_n)$ is the joint PDF between two target ranges. Typically, one can evaluate the integral for uniform distribution for the two target ranges. When $\mathbf{R}_{n}$ is a diagonal matrix, this function simplifies to
\begin{align}
\mathcal{A}_r(r_m,r_n) = \frac{1}{\sigma_n^2}  \mathbb{E} \left( \left[  \mathbf{x}_n  \circ \mathbf{t}_n \right]^T \left[  \mathbf{x}_m  \circ \mathbf{t}_m \right]  \right),
\end{align}
which measures the ambiguity or dissimilarity between the time weighted columns of $\mathbf{X}_i$. We refer this function as \emph{target ambiguity function} (TAF) which is different from the ambiguity function known in radar literature. TAF measures the similarity of the received vectors due to each target rather than the similarity between the ideal and time-frequency shifted transmit waveform.  Moreover, TAF operates on the derivative of the vector signal received from each target rather than the received signal vector. Even if the reflection from the targets are correlated $\mathbb{E}[\gamma_m \gamma_n] \ne 0$, one can design waveform and array geometry which ensures TAF to be very small between two targets, its parameters can be estimated more accurately, as observed in the sequel.  

From \eqref{eqn:eqn18}, the expected value of second order partial derivative for range can be written as in \eqref{eqn:pd_range0}, which can be the inverse of the MSE if off-diagonal elements of the FIM are zero.\footnote{In general, the FIM is block diagonal in nature, where the 3x3 block matrices corresponding to all parameters for a given target can influence each other on the MSE. However, the parameters are one target does not influence the MSE of another. Also note that inverse of block diagonal matrix is a block diagonal matrix with the individual inverse of the blocks.} It can be noticed that MSE is inversely proportional to the SNR, RCS parameter and norm of the derivative of the signal vector from that target. Hence, one should design a transmitter waveform which maximizes the average norm, $\mathbb{E} \left( \|  \mathbf{x}_m  \circ \mathbf{t}_m \|^2 \right)$ rather than the waveform with best self-correlation property. At the same time, one also wishes to minimize $\mathbb{E} \left( \left(\mathbf{x}_m  \circ \mathbf{t}_m\right)^T \left(\mathbf{x}_n  \circ \mathbf{t}_n\right) \right)$, to reduce the impact due to cross terms in the FIM. 

\subsubsection{TAF in Velocity}
Similar to the previous case, one can find the partial derivatives with respect to the velocity parameter as shown below. Note that,
\begin{align}
& \frac{\partial\ x(\mathbb{T}_m[t-i T_p - 2 \tau_m - \varphi_l])}{\partial\ v_m} \nonumber \\
& = -2 \mathbb{T}_m[t-i T_p - 2 \tau_m - \varphi_l] \left(\frac{2c}{(c-v_m)^2} \cdot \frac{\pi^2}{\beta^2} \right) \nonumber \\
& ~~~~~~~~ x(\mathbb{T}_m[t-i T_p - 2 \tau_m - \varphi_l]) \circ \mathbf{t}_m \circ \mathbf{t}_m,
\end{align}
using the fact that $\frac{d \mathbb{T}_m}{dv} = \frac{2c}{(c-v_m)^2}$. We can write the second order derivative with respect to the velocity parameter as given in \eqref{eqn:pd_velocity0}. Thus, for minimizing the MSE for velocity estimation, one should design waveform which maximizes the norm of the second order derivative of the transmitter waveform. Moreover, define the TAF for velocity as
\begin{equation}
\mathcal{A}_{v}(v_m, v_n) = \mathbb{E} \left( \left[  \mathbf{x}_n  \circ \mathbf{t}_n^2 \right]^T \mathbf{R}_{n}^{-1} \left[  \mathbf{x}_n  \circ \mathbf{t}_n^2 \right] \right) .
\label{eqn:TAF_vel}
\end{equation}
As in the case of range, the above TAF is computed over all possible velocity values except when $v_m = v_n$. 

\subsubsection{TAF in Azimuth}
Evaluation of the partial derivatives with respect to the azimuth angle gives the TAF in angle domain. From the definition of $\varphi_l$ , we can write
\begin{align}
& \varphi_l  
%=  \frac{\mathbf{p}_m^T \mathbf{p}_l}{c} \nonumber \\
 ~~~ =  \frac{\cos \phi_m \cos \theta_m x_l \hspace{-0.1cm} + \hspace{-0.1cm} \sin \phi_m \cos \theta_m y_l \hspace{-0.1cm} + \hspace{-0.1cm} \sin \theta_m z_l}{c}, 
\end{align}
where $(x_l, y_l, z_l)$ are the Cartesian coordinates of the $l^{th}$ receiver antenna, $\phi_m$ is azimuth angle of the $m^{th}$ target and $\theta_m$ is the corresponding elevation angle of the same target. Now, one can write the second order partial derivative as given in \eqref{eqn:pd_angle0}, and the TAF in angle can be written as 
\begin{equation}
\mathcal{A}_{\phi}(\phi_m, \phi_n) \hspace{-0.1cm} = \hspace{-0.1cm} \mathbb{E} \left( \left[  \mathbf{x}_n \hspace{-0.05cm} \circ \hspace{-0.05cm} \mathbf{t}_n \hspace{-0.05cm} \circ \hspace{-0.05cm} \mathbf{\Phi}_n \right]^T \hspace{-0.05cm} \mathbf{R}_{n}^{-1} \hspace{-0.05cm} \left[  \mathbf{x}_m \hspace{-0.05cm} \circ \hspace{-0.05cm} \mathbf{t}_m \hspace{-0.05cm} \circ \hspace{-0.05cm} \mathbf{\Phi}_m  \right] \right) ,
\label{eqn:TAF_ang}
\end{equation}
where $\mathbf{\Phi}_m$ is defined as the vector of partial derivatives of $\varphi_l$ with respect to the angle $\phi_m$ for all $l=1,\ldots,L$. That is,
\begin{align}
& \mathbf{\Phi}_m 
%\triangleq \left\{ \frac{\partial \varphi_l}{\partial \phi_m} \right\}, \nonumber \\
 = \frac{\left\{ \sin \theta_m \left(y_l \cos \phi_m - x_l \sin \phi_m \right) + z_l \cos \theta_m \right\}}{c}.
\end{align}
Thus, this ambiguity function brings in the role of the array geometry to improve the MSE bound by minimizing the TAF. 

\subsubsection{Cross Terms in FIM}
Using the above expressions, one can write the expression for the mixed second order derivatives for pairs of terms such as (range, velocity) or (range, angle) and (velocity, angle). Since these terms are not zero, the FIM has the structure of a block diagonal matrix, where each block belongs to one target. That is, under independent scattering assumption, one target does not influence the parameter estimation of the other target provided we design waveforms which minimizes the TAFs. Hence, to achieve the bounds given in Sec.~\ref{sec:EstBounds}, we need to minimize all inner products given in \eqref{eqn:crossTerms}, so that the non-diagonal terms in FIM are reduced.

\begin{figure*}
\begin{align}
& \mathbb{E}\left[\frac{\partial^2 \log f_z(\mathbf{z}_i)}{\partial r_m v_m} \right]  = - \frac{16 \rhot \pi^4 \mathbf{T}_m^3 \mathbb{E}[\gamma_m^2] \mathbb{E}[\|\zeta_{m}\|^2] \epsilon^4 r_m^{-4e} }{\beta^4 (c-v_m)^2}  \left( \mathbf{x}_m  \circ \mathbf{t}_m \circ \mathbf{t}_m \right)^T  \left( \mathbf{x}_m  \circ \mathbf{t}_m \right) \nonumber \\
& \mathbb{E}\left[\frac{\partial^2\ \log f_z(\mathbf{z}_i)}{\partial\ r_m \phi_m} \right]  =  - \frac{8 \rhot \pi^4 \mathbf{T}_m^3 \mathbb{E}[\gamma_m^2] \mathbb{E}[\|\zeta_{m}\|^2] \epsilon^4 r_m^{-4e} }{\beta^4 \ c}  \left( \mathbf{x}_m  \circ \mathbf{t}_m \circ \mathbf{\Phi}_m \right)^T   \left( \mathbf{x}_m  \circ \mathbf{t}_m\right) \nonumber \\
& \mathbb{E}\left[\frac{\partial^2\ \log f_z(\mathbf{z}_i)}{\partial\ v_m \phi_m} \right]  = - \frac{8 \rhot \pi^4 \mathbf{T}_m^2 \mathbb{E}[\gamma_m^2] \mathbb{E}[\|\zeta_{m}\|^2]\ c \epsilon^4 r_m^{-4e}}{\beta^4 \ (c-v_m)^2}  \left( \mathbf{x}_m  \circ \mathbf{t}_m \circ \mathbf{\Phi}_m \right)^T   \left( \mathbf{x}_m  \circ \mathbf{t}_m \circ \mathbf{t}_m\right) 
\label{eqn:crossTerms}
\end{align}
\hrulefill
\end{figure*}

\subsubsection{Generalized Target Ambiguity Function}
Although we have listed the ambiguity functions that arose from the partial derivatives taken with respect to different parameters, one can combine all to define a generalized TAF from which other TAFs can be derived as special cases. For instance, the generalized TAF given in \eqref{eqn:genTAF} includes variations in range, velocity and array geometry.

\begin{figure*}
\begin{align}
& \mathcal{A}(r_1,r_2,v_1,v_2,\phi_1,\phi_2) \hspace{-0.05cm} = \hspace{-0.05cm} \mathbb{E}\left[\mathbf{x}_m(r_1,v_1,\phi_1) \circ \mathbf{t}_m(r_1,v_1,\phi_1) \hspace{-0.05cm} \circ \hspace{-0.05cm}  \mathbf{\Phi}_m(\phi_1) \right]^T \mathbf{R}_{n}^{-1}        \left[\mathbf{x}_m(r_2,v_2,\phi_2) \hspace{-0.05cm} \circ \hspace{-0.05cm} \mathbf{t}_m(r_2,v_2,\phi_2) \hspace{-0.05cm} \circ \hspace{-0.05cm} \mathbf{\Phi}_m(\phi_2) \right] \hspace{-0.05cm} .
    \label{eqn:genTAF}
\end{align}
\hrulefill
\end{figure*}

\textcolor{black}{
\subsection{Proof of Theorem 1} \label{app:cov}
Consider the cost in \eqref{eqn:CostFunc} without scaling factor $1/E$, whose partial derivative with respect to $y$ can be written as:
\begin{equation}
\frac{\partial F(y, \dot{y})}{\partial y} \hspace{-0.1cm} = \hspace{-0.1cm} \int_{-\frac{T_s}{2}}^{\frac{T_s}{2}} \left[ \dot{y}(t) \frac{ \partial \dot{y}(t+\delta)}{\partial y} \hspace{-0.1cm} + \hspace{-0.1cm} \frac{ \partial \dot{y}(t)}{\partial y} \dot{y}(t+\delta) \right]  dt, 
\label{eqn:lhs38} 
\end{equation}
where $\frac{ \partial \dot{y}(t)}{\partial f}$ denotes the variation in the temporal slope of $y(t)$ due to the functional variation in $y$. Similarly,
\begin{eqnarray}
\frac{\partial F(y, \dot{y})}{\partial \dot{y}} 
& = & \int_{-\frac{T_s}{2}}^{\frac{T_s}{2}} \left[ \dot{y}(t)  +  \dot{y}(t+\delta) \right]   dt = 0,
\end{eqnarray}
for $y(t)$ whose end values are equal. 
The right hand side of \eqref{eqn:EulerEqn} can be written as
\begin{equation}
\frac{d}{dt} \left( \frac{\partial F(y, \dot{y})}{\partial \dot{y}} \right)  =  0.
\label{eqn:rhs38} 
\end{equation}
%\left[ \dot{y}(t)  +  \dot{y}(t+\delta) \right]
Upon equating \eqref{eqn:lhs38} and \eqref{eqn:rhs38}, we get the conditions for optimality as
\begin{equation}
\int_{-\frac{T_s}{2}}^{\frac{T_s}{2}} \left[ \dot{y}(t) \frac{ \partial \dot{y}(t \hspace{-0.1cm} + \hspace{-0.1cm} \delta)}{\partial y} \right] dt \hspace{-0.1cm} = \hspace{-0.1cm} - \int_{-\frac{T_s}{2}}^{\frac{T_s}{2}} \left[ \frac{ \partial \dot{y}(t)}{\partial y} \dot{y}(t \hspace{-0.1cm} + \hspace{-0.1cm} \delta) \right]   dt.
\label{eqn:FirstVar0}
\end{equation}
Consider the first variation of the function as
\begin{equation} 
y^{n+1}(t) = y^{n}(t) \pm \lambda~ y(\alpha~t + \beta),
\label{eqn:FirstVar1}
\end{equation} 
where $\lambda > 0$, $\alpha > 0$ and $\beta \in \left[-\frac{T_s}{2},\frac{T_s}{2}\right)$ are constants. The number of time shifts $\beta$ are finite, since the number of dimensions of the considered functions -- which is $2 BW T_s$ -- in $\mathcal{L}_2$ are finite. Therefore, $\beta$ will be changed in steps $\frac{1}{2 BW}$ to accommodate the Nyquist sampling rate. Similarly, choice of $\alpha$ also is limited within the range $\left[\frac{1}{2 BW T_s},1 \right)$. For various choices of $\beta$ and amplitude scale $\lambda$, $y(t)$ is modified 
as given in \eqref{eqn:FirstVar1}. Now, the optimality condition can be evaluated as follows. Let $z(t)= y(\alpha~t-\beta)$. Then,
\begin{eqnarray}
y_1(t) & = & y(t) \left( 1+\frac{\lambda~ z(t)}{y(t)}\right), \nonumber
\end{eqnarray}
\begin{eqnarray}
\frac{\partial y_t(t)}{\partial y(t)} & = & \left(1 + \frac{\lambda~z}{y} \right) + y \left( \frac{-\alpha~\lambda~z}{y^2}\right) \left( \frac{\partial z}{\partial y}\right)  \nonumber \\
%& = & \left(1 + \frac{c~z}{y} \right) - \left( \frac{\alpha~\lambda~z}{y}\right) \left( \frac{\alpha} { \dot{y}}\right), \\
& = & \left(1 + \frac{\lambda~z}{y} \right) - \alpha^2 \left( \frac{\lambda~z}{y \dot{y}}\right), 
\label{eqn:FirstVar2}
\end{eqnarray}
where the time variable is dropped for brevity only positive sign is considered in \eqref{eqn:FirstVar1}, without loss of generality. Moreover, we have used the fact that 
\[
\frac{\partial z}{ \partial y} = \frac{\partial y(\alpha t-\beta)}{\partial y} = \alpha \frac{\partial t}{ \partial y} = \left( \frac{\alpha}{\dot{y}} \right) .
\]
By substituting \eqref{eqn:FirstVar2} in \eqref{eqn:FirstVar0}, and using some algebra, one can write the following where the time dependence is notated as a subscript. 
\begin{equation}
\frac{\dot{y}_t}{\dot{y}_{t+\delta}}~\left(1 -  
\frac{\alpha^2}{\dot{y}_{t+\delta}} \right) =  \frac{y_{t+\delta}}{y_{t}} \left(1  -  \frac{\alpha^2}{\dot{y}_t} \right) .
\label{eqn:optimumCond1} 
\end{equation}
That is, the above condition is satisfed if the slope of the signal does not change drastically for small delta, i.e.~if $y(t)$ is continuous, since $y_t \approx y_{t+\delta}$ and $\dot{y}_t \approx \dot{y}_{t+\delta}$ and the approximation error can be evaluated using the Taylor series expansion. 
%That is,
%\begin{eqnarray}
%\frac{y_{t+\delta}}{y_t} & = & 1 + %\left(\frac{\delta}{1!}\right) %\frac{\dot{y}_t}{y_t} + %\left(\frac{\delta^2}{2!}\right) %\frac{\ddot{y}_t}{y_t} + \ldots \nonumber \\ 
%\frac{\dot{y}_{t+\delta}}{\dot{y}_t} & = & 1 + %\left(\frac{\delta}{1!}\right) %\frac{\ddot{y}_t}{\dot{y}_t} + %\left(\frac{\delta^2}{2!}\right) \frac{d~ %\ddot{y}_t}{dt}\frac{1}{\dot{y}_t} + \ldots %\nonumber 
%\end{eqnarray} 
By substituting the first order approximations for the two ratios $\frac{y_{t+\delta}}{y_t}$ and $\frac{\dot{y}_{t+\delta}}{\dot{y}_t}$ in \eqref{eqn:optimumCond1}, we get a quadratic equation in $\delta$. Evaluating the condition for roots of this equation to be real, gives the required result.
%\begin{equation}
%\underbrace{\frac{\left(1 %-\frac{\alpha^2}{\dot{y}_{t+\delta}} %\right)}{\left(1  -  \frac{\alpha^2}{\dot{y}_t} %\right)}}_{=\Gamma} \approx   
%\left[1 + \delta \frac{\dot{y}_t}{y_t} \right] 
%\left[1 + \delta \frac{\ddot{y}_t}{\dot{y}_t} %\right] 
%\label{eqn:optimumCond2} 
%\end{equation}
%By expanding the RHS in the above equation into a %quadratic equation in $\delta$, one can find the %conditions in which $\delta$ will result in real %solution. That is,
% \begin{eqnarray}
% \left(\frac{\dot{y}_t}{y_t} + \frac{\ddot{y}_t}{\dot{y}_t}\right)^2 + 4 \left( \frac{\ddot{y}_t}{y_t}\right) \left[1 - \frac{\left(1 -\frac{\alpha^2}{\dot{y}_{t+\delta}} \right)}{\left(1  -  \frac{\alpha^2}{\dot{y}_t} \right)} \right] \ge 0
% \end{eqnarray}
%where we have used the approximation,
%\begin{equation}
%4 \left( \frac{\ddot{y}_t}{y_t}\right) \left[1 - \Gamma \right] \approx \frac{4 \delta \left(\dot{y}_{t+\delta}-\dot{y}_t \right)^2}{y_t \dot{y}_{t+\delta} \left(\alpha^2  - \dot{y}_t \right)},
%\end{equation}
%based on the approximation $\ddot{y}_t \approx \delta \left(\dot{y}_{t+\delta} - \dot{y}_t \right) $. 
%Now, one need to choose the value of $\alpha$ which meets the above constraints for all values of $t$ for sufficiently small value of $\delta$. 
%Thus proved. 
}
%If the solution is not obtained, reduce the value of $\delta$ and com pute the value of $\alpha$. If there are multiple values of $\alpha$ which meets the constraints, we need to try all the values of $\alpha$ and choose the one for which maximum cost function reduction is achieved.  For the chosen $\alpha$ value, try various values of $\beta$ for which the cost function is minimized by the maximum extend. This iterative procedure can be used to obtain the optimum waveform $y(t)$ which minimises the cost function $F(y, \dot{y})$. 

\begin{figure*}[!t]
\centering
\subfigure[]{
    \includegraphics[width=0.31\textwidth]{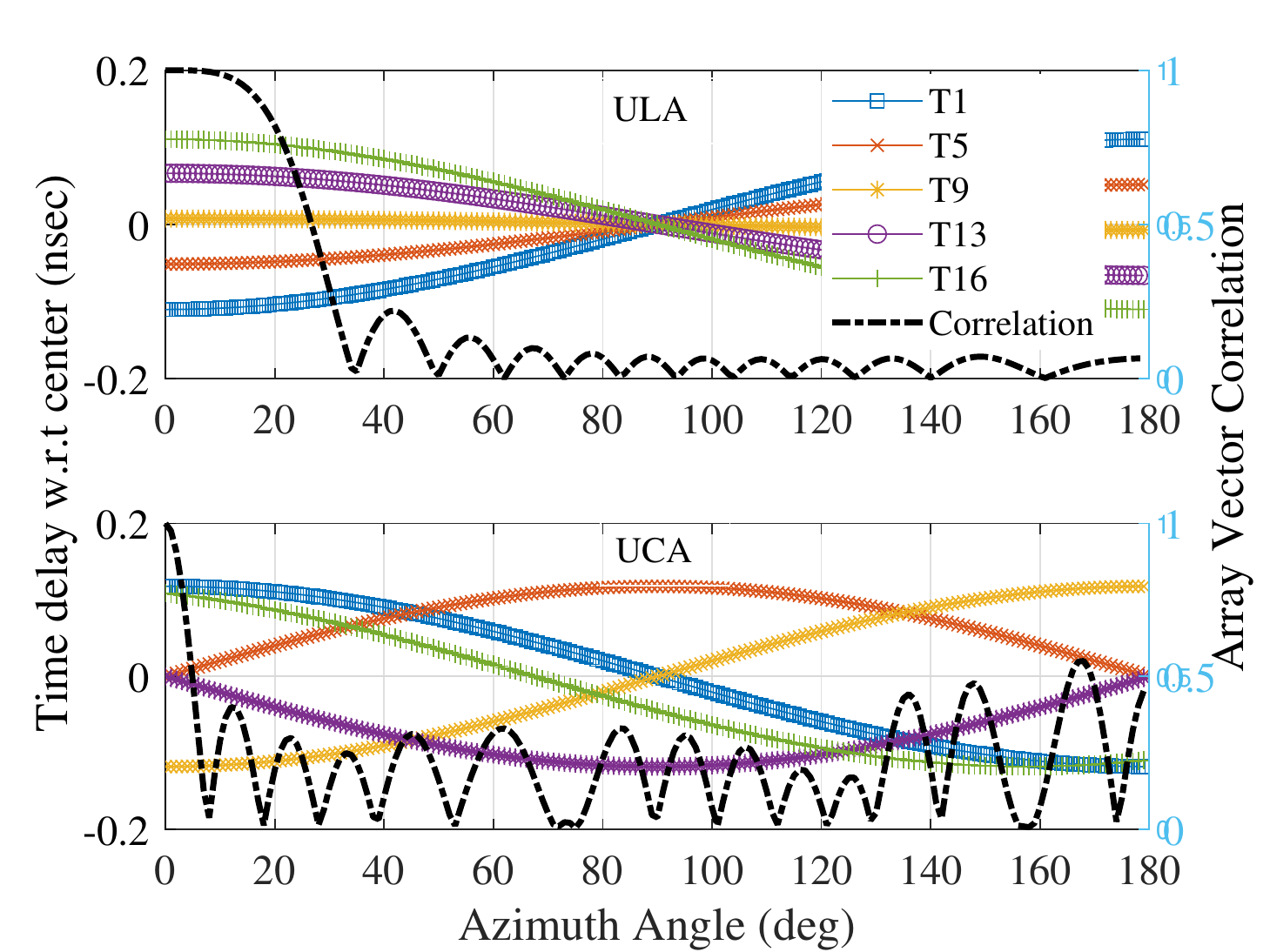}
    \label{fig:ula_corr}
}
\subfigure[]{
    \includegraphics[width=0.31\textwidth]{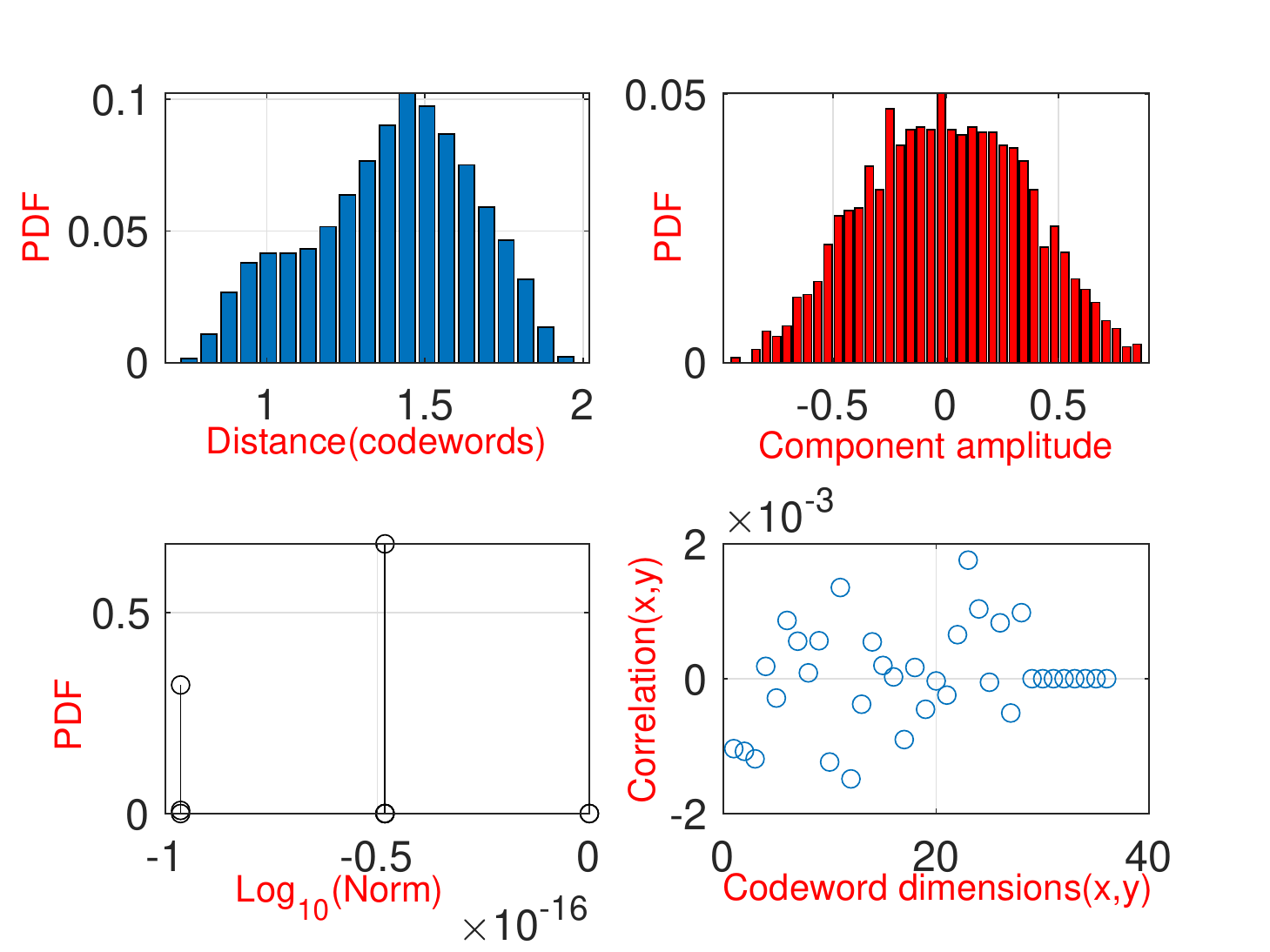}
    \label{fig:codeWordStat}
}
\subfigure[]{
    \includegraphics[width=0.31\textwidth]{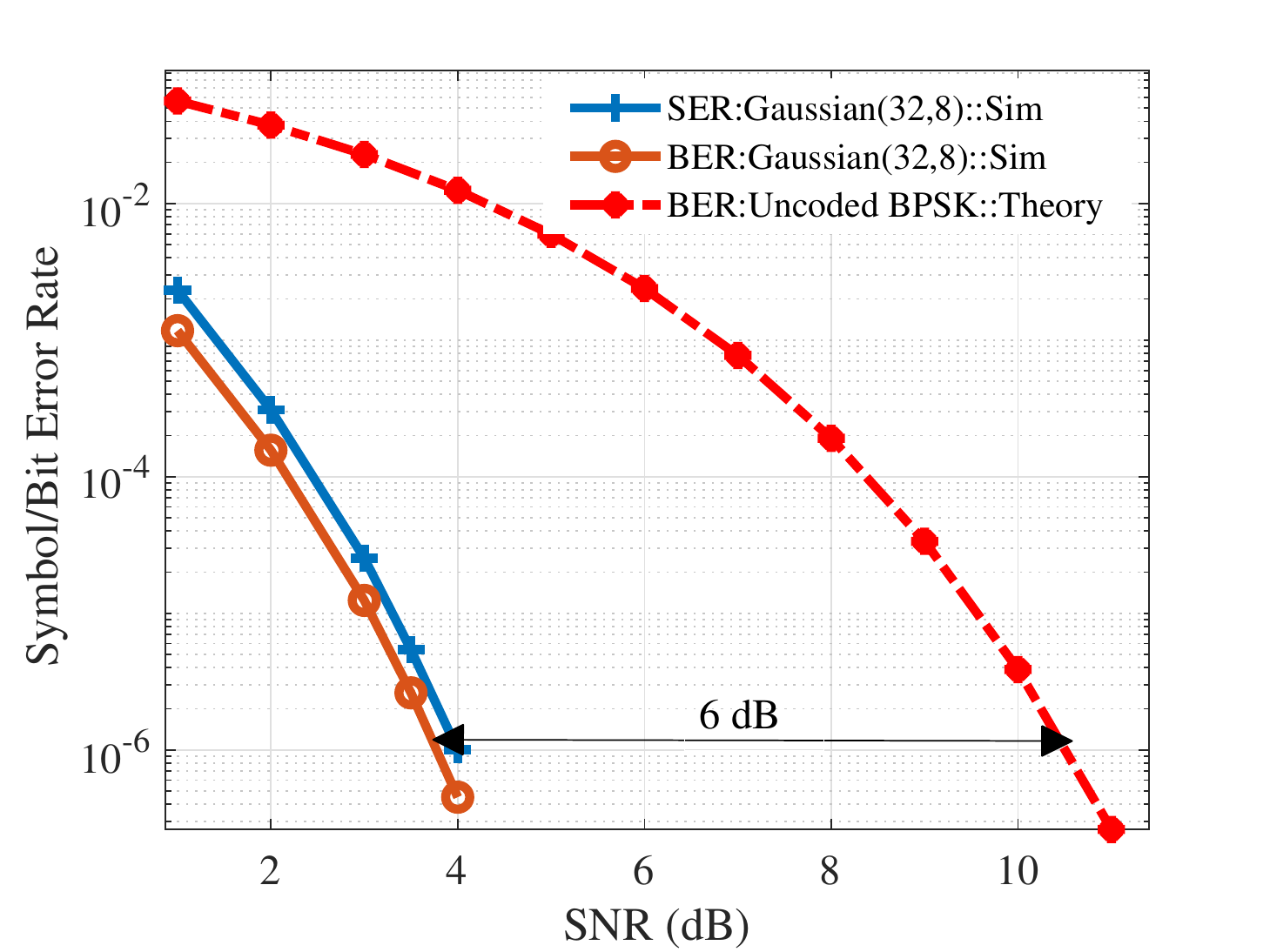}
    \label{fig:codeWordBER}
}
\vspace{-0.3cm}
\caption{\textcolor{black}{(a) Time delay across antenna elements with elevation angle $45^\circ$ for ULA and UCA with $L=12$, (b) Code book Statistics for $(32,8)$ Gaussian spherical code, and (c) Bit and symbol error rate performance of Gaussian spherical code $(32,8)$ in an AWGN channel.}}
\label{fig:subfigureExample2}
\hrulefill
\end{figure*}

\textcolor{black}{
\subsection{Array Vector Correlation}
\label{app:arrayCorr}
Here, we study the correlation between two array steering vectors for two different array geometries at various steering angles. Consider a uniform linear array (ULA) and a uniform circular array (UCA) with $16$ elements. The parameters listed in Table~\ref{tab:antParams} were used for realizing the two array geometries.
}

\begin{table}[htb]
\begin{center}
\begin{tabular}{|c|l|c|c|} 
\hline 
Sl. & Parameter & ULA & UCA \\ 
\hline \hline
1 & Element spacing & $\frac{\lambda_c}{2}$ &   $\pi~\frac{\lambda_c}{2}$ \\ 
\hline 
2 & Array diameter & $7.5~\lambda_c$ &  $8~\lambda_c$ \\ 
\hline 
3 & Origin & at the center & at the center \\ 
\hline 
4 & Orientation & along $X-$axis & along $X-Y$ plane \\
\hline
\end{tabular} 
\caption{Parameters for ULA and UCA with $\lambda_c=12.5$ mm and $L=16$.}
\label{tab:antParams}
\end{center}
\end{table}

\textcolor{black}{
Figure~\ref{fig:ula_corr} shows the time delays across antenna elements with respect to the origin for ULA and UCA respectively, for various azimuth angles while the elevation angle of the source is kept at $45^\circ$. The plots also show the correlation among the array steering vectors. Note that UCA shows a higher correlation for angles outside the main beam direction, compared to the ULA. However, the main beam width for UCA is smaller than that of ULA, for a given $L$. These time delay variations also 
%for ULA and UCAdemonstrate another important parameter which will define the various sampling time instants at which the received pulse will be sampled before it will be correlated with the transmit pulse.
influence the TAFs, through the sampling time instants for the received pulse before correlating it with the transmit pulse. 
%Hence, they define the TAFs. 
For various angles, the time delays in UCA are not concentrated around zero as opposed to ULA. This causes wider peaks in the TAFs. That is, for azimuth angles close to $90^\circ$ degrees, all elements receive the data nearly at the same time and the time differences are nearly zero. Hence, UCA will be preferred array geometry for obtaining good $SB_{\text{angle}}$ for the given number of antenna elements. }
% \begin{figure}[!thb]  
%   \begin{center}
%  % \vspace{-1.5in}
%   \includegraphics[width=0.5\textwidth]{ula_uca_array_corr.pdf}
%  \end{center}
%  % \vspace{-1.5in}
% \caption{Time Delay across elements with respect to the origin and Correlation among steering vectors for various azimuth angles while elevation angle is kept at 45 degrees for ULA and UCA with 12 antenna elements.}
% \label{fig:ula_corr}
% \end{figure}

%\begin{figure}[!thb]  
%  \begin{center}
% \vspace{-1.5in}
%  \includegraphics[width=3.5in]{ci%rcArrayCorr.pdf}
% \end{center}
%  \vspace{-1.5in}
%\caption{UCA: Time Delay across elements with respect to the origin and Correlation among steering vectors for various azimuth angles while elevation angle is kept at 45 degrees.}
%\label{fig:uca_corr}
%\end{figure}

\subsection{Code Construction Procedure} 
\label{app:codeConstr}
It is well-known that Gaussian random source achieves the capacity in a Gaussian channel. Hence, the code construction for the data only transmission can adopt any of the well known methods such as the lattice Gaussian coding \cite{Ling_TIT_2014}, \cite{Shannon_BSTJ_1959}. We elaborate a practical code construction along the lines of Shannon's spherical code idea. The $N$-dimensional Gaussian vector with components, from i.i.d. Gaussian random variables with zero mean and unit variance, can be scaled by $\sigma_x$ to obtain the desired codeword with average transmitter power $P=\sigma_x^2$. A Gaussian spherical code with desired number of codewords, e.g.~$2^r$ codewords or $r$ bits per codeword, can be constructed by partitioning the surface area of an $N$-dimensional sphere with unit radius. This can be done by well known methods such as K-means algorithm \cite{Gray_SPMAG_1984}. Now, use the \textcolor{black}{normalized} centroids of the $2^r$ regions as the desired codewords for data transmission. Such a code is known to achieve the capacity of an AWGN channel for large $N$ \cite{Shannon_BSTJ_1959}.

Figure~\ref{fig:codeWordStat} shows the properties of one of such codes designed for length $N=32$, code rate $r=1/4$. K-means algorithm was used to design $256$ size codebook with $32$ dimensions per codeword. $10^5$ random Gaussian vectors were used as input to the K-means algorithm after they were normalized to have unit norm. About $200$ iterations were performed so that the update error between two iterations is less than $10^{-5}$. It can be noticed that the all the codewords have unit norm, and the distance between them is between $1$ and $2$. The correlation between the various dimensions is nearly zero, in the order of $10^{-3}$. The PDF of the components of the codewords closely approximate the Gaussian distribution. Figure~\ref{fig:codeWordBER} demonstrates the bit error rate and symbol error rate performances in an AWGN channel with a maximum likelihood decoder, where a coding gain of $6$ dB is observed.
%where the BER values match the  
%Bit error rate (BER) for this code in AWGN channel can be simulated easily and simulation shows that this code achieves the 6 dB coding gain, as predicted by theory.  
%theoretical computed values for the given $SNR = \frac{E_b}{N_0}$ values. 

% \begin{figure}[!thb]  
%   \begin{center}
%   %\vspace{-1.2in}
%   \includegraphics[width=3.5in]{PSR_codeBook_params.pdf}
%  \end{center}
%   %\vspace{-1.2in}
% \caption{Code book Statistics for $(32,8)$ Gaussian spherical code.}
% \label{fig:codeWordStat}
% \end{figure}

%\begin{figure}[!thb]  
%  \begin{center}
%  \vspace{-1.5in}
%  \includegraphics[width=3in]{PSR_codeBook32_BER.pdf}
% \end{center}
%  \vspace{-1.5in}
%\caption{BER and SER Performance of Gaussian Spherical Code $(32,8)$ in AWGN Channel.}
%\label{fig:codeWordBER}
%\end{figure}

\end{appendix}

\balance

\bibliographystyle{IEEEtran}
\bibliography{IEEEabrv,RITRefs}

\end{document}